\documentclass[10pt,journal,compsoc]{IEEEtran}

\ifCLASSOPTIONcompsoc
\usepackage[nocompress]{cite}
\else
\usepackage{cite}
\fi
\usepackage{enumerate}
\usepackage{graphicx}
\usepackage{amsmath}
\usepackage{bm}
\usepackage{mathrsfs}
\usepackage{algorithm}
\usepackage{algorithmic}
\usepackage{subcaption}
\usepackage[justification=centering]{caption}
\usepackage[dvipsnames]{xcolor}
\usepackage{gensymb}
\usepackage{multirow}

\ifCLASSINFOpdf
\else
\fi
\begin{document}
		%
		\title{Vital Sign Monitoring in Dynamic Environment via mmWave Radar and Camera Fusion}
		%
		%
		%
		%
		
		\author{Yingqi~Wang,~\IEEEmembership{Student Member,~IEEE,}
			Zhongqin Wang,
			J. Andrew Zhang, \IEEEmembership{Senior Member, IEEE},
			Haimin Zhang,
			and Min Xu, \IEEEmembership{Member, IEEE}
			\IEEEcompsocitemizethanks{
				\IEEEcompsocthanksitem Yingqi Wang, Zhongqin Wang, J. Andrew Zhang, Haimin Zhang and Min Xu (Corresponding Author) are with the School of Electrical and Data Engineering, University of Technology Sydney, Sydney 2007, Australia.
				E-mail: \{yingqi.wang\}@student.uts.edu.au
				E-mail: \{zhongqin.wang, andrew.zhang, haimin.zhang, min.xu\}@uts.edu.au\protect\\}}

	\IEEEtitleabstractindextext{%
		\begin{abstract}
			Contact-free vital sign monitoring, which uses wireless signals for recognizing human vital signs (i.e, breath and heartbeat), is an attractive solution to health and security. However, the subject's body movement and the change in actual environments can result in inaccurate frequency estimation of heartbeat and respiratory. In this paper, we propose a robust mmWave radar and camera fusion system for monitoring vital signs, which can perform consistently well in dynamic scenarios, e.g., when some people move around the subject to be tracked, or a subject waves his/her arms and marches on the spot. Three major processing modules are developed in the system, to enable robust sensing. Firstly, we utilize a camera to assist a mmWave radar to accurately localize the subjects of interest. Secondly, we exploit the calculated subject position to form transmitting and receiving beamformers, which can improve the reflected power from the targets and weaken the impact of dynamic interference. Thirdly, we propose a weighted multi-channel Variational Mode Decomposition (WMC-VMD) algorithm to separate the weak vital sign signals from the dynamic ones due to subject's body movement. Experimental results show that, the 90${^{th}}$ percentile errors in respiration rate (RR) and heartbeat rate (HR) are less than 0.5 RPM (respirations per minute) and 6 BPM (beats per minute), respectively.
		\end{abstract}
		
		}

	\maketitle

	\IEEEdisplaynontitleabstractindextext

	%
	\IEEEpeerreviewmaketitle

	\IEEEraisesectionheading{\section{Introduction}\label{sec:introduction}}

	%
	%
	%
	%
	
	\IEEEPARstart{R}{ecent} years have witnessed the rapid development of vital sign monitoring. Under the COVID-19 crisis, the demand for vital sign monitoring systems for tracking patients' health conditions becomes more urgent. Traditional vital sign measurements based on PPG and ECG sensors require wearing electrodes or chest bands, which are uncomfortable and inconvenient for some users like infants or burned patients. In this case, the technology of contact-free vital sign monitoring has been indispensable for today's healthcare systems.
	
	Currently, Radio frequency (RF) based sensing methods have been widely investigated in contact-free vital sign monitoring\cite{wifi0, c10, cotsmmwave0, rfid3}. It is achieved by capturing the tiny shift from the chest caused by heartbeat and respiratory movements. \textit{(1) WiFi.} WiFi-based monitoring has been explored due to its easy availability and cost-saving infrastructures\cite{c1, c2}. Some works\cite{c4, c5, wifi2, wifi3} apply the channel state information (CSI) to sense the chest movement since the CSI contains the phase information, which could provide a much higher sensing resolution than the received signal strength (RSS). However, most commercial off-the-shelf (COTS) WiFi devices cannot provide high enough bandwidth to separate multi-path noise in different range dimensions. Also, the transmitter and receiver are required to be deployed separately. \textit{(2) Portable COTS mmWave radar.} Currently, the COTS mmWave radars\cite{cotsmmwave1} have also received increasing attention. They are typically developed based on the FMCW technique and can achieve higher range and velocity resolutions. Furthermore, they enable us to locate and distinguish different targets of interest. In this case, it is becoming the mainstream choice in the field of multi-target vital sign detection. However, most of the existing works\cite{letIOT, mmecg, Enhancement77} ignore the multipath and dynamic interference caused by the environment and moving persons in actual deployments and setup in an ideal environment. In fact, daily life sensing scenarios could be more complicated. For example, for indoor healthcare, the users may stay at a spot such as sofa, desk or table for a long time and perform activities such as computer work, exercise and entertainment, while other family members may walk past them in the room from time to time. Such complicated environment brings three challenges in contact-free vital sign monitoring, which have not been thoroughly addressed yet in the literature.
	
	
	\textit{1)} Accurate localization of static person of interest in dynamic scenarios is challenging. In practice, there are various static objects in different places. All of these will reflect the signals generated by RF device, causing the multipath effect. The multipath signal will result in a fake person object. It is very challenging to identify which one is the real target without any prior knowledge. Consequently, the range-bin data from the COTS mmWave radars may be wrongly chosen for vital sign recognition.
	
	\textit{2)} The movements of nearby people in space can cause interference. When other persons near to the person to be monitored are moving, they can cause strong reflections to disturb the localization of the target and bury the slight movement of breathing and heartbeat. The conventional range-doppler \cite{mmVib} highlights dynamic objects via the velocity. But it could become less efficient for separating location-fixed targets from static background objects, as both have near zero doppler frequencies. We call it as range overlapping phenomenon in this paper. Estimating vital signs from this condition is challenging.
	
	\textit{3)} The motion of the target may severely interfere with the vital sign estimation. In an actual environment, even if a subject is located on the same spots or sits on a chair, he/she cannot keep entirely static\cite{c19, c20}. The target might move the body parts such as arms, torsos and limbs at any time. The COTS mmWave radar cannot provide a high enough angle resolution, so it is difficult to identify the precise position of the chest for recognition. As a result, the vital sign signals will be blocked by these body-motion noise.
	
	In this paper, we propose a hybrid multi-objective vital sign monitoring scheme, called Vision-assisted radar Vital Sign Monitor (VaR-VSM). This scheme is capable of monitoring vital signs for the location-fixed human target stably in the presence of interferences from both body movements of the target and the movement of other people. In other words, our scheme can achieve reliable vital sign detection for location-fixed target in a multi-people environment, where the target may have body parts movement and people not being monitored can move around. It firstly fuses a camera and COTS mmWave radar to localize the person of interest. Once the positions of targets are identified, the beamforming techniques are simultaneously performed on Tx and Rx antennas to improve the signal-to-noise ratio (SNR) of received echoes from the targets. To suppress the effect of the subject's body motion, we propose weighted-multi-channel Variational Mode Decomposition (WMC-VMD) which is an enhanced version of VMD\cite{c36}. Although our scheme is designed and optimized primarily for sensing location-fixed targets, it is also capable of sensing moving targets and can still work in partially blocked transmission (with light obstacles between radar and targets). Our main contributions are summarized below:
	
	\textit{1)} We introduce a method using the object detection algorithm in computer vision to detect all people and locate the target in the scenario. We then use the COTS mmWave radar to generate the angle-of-arrival (AoA) heatmap in the range and angle dimensions. On this basis, we map the detected image objects to the AoA heatmap for accurate target localization. Compared to the pure mmWave localization methods\cite{cotsmmwave1, c35}, the proposed fusion scheme could significantly reduce the false positive error.
	
	\textit{2)} We propose to use real-time transmit and receive beamforming to enhance the energy of the radar echoes from the targets and reduce the interference from interfering objects. Leveraging the calculated position information, we apply the Transmitting beamforming (Tx-BF) and Receiving beamforming (Rx-BF) and control the main lobe of the transmitting antennas toward certain directions. We also calculate the position-based beamforming weights to further suppress the effect of noise in unwanted directions. By combining the two beamforming methods, we can effectively improve the SNR and reduce the interference from nearby persons.
	
	\textit{3)} We propose the WMC-VMD method to reduce the effect of the body motion of the target. The method can adaptively leverage the information of multiple channels from range bins to improve the decomposition efficiency and suppress the motion impact caused by body motion. A strategy to accelerate the computation of WMC-VMD is also proposed to fulfill real-time application.

	\textit{4)} We build a prototype to validate the proposed system VMVSM. Experimental results show that in the actual interference environment, the 90${^{th}}$ percentile errors in respiratory and heartbeat rates are less than 0.5 RPM and 6 BPM, respectively. The estimation accuracy is much higher than what can be achieved by state-of-the-art technologies.

	\section{Related Work}
	Some existing works have been conducted to tackle the above-mentioned challenges. The work \cite{c21} employs different radar layouts to detect breath during sleeping scenarios when body motion happens. Nonetheless, it requires to manually adjust the sensor position according to different gestures. Another work\cite{c22} based on the rotation of Rx and Tx antennas can detect posture change whereas the extra mechanical rotator is required. Vital-Radio \cite{c23} considers that the limb motion pattern is aperiodic and uses this property to filter the vital sign signal. WiSpiro\cite{c24} trains a neural network to recognize the body motion to navigate the radar to move to the front of the target. But it requires a large rail to place the system.
	
	
	Despite that no work has fully resolved the second challenge, there are still some inspiring works that have demonstrated the potential. Independent component analysis (ICA) based algorithms have been investigated in the works\cite{c28, c29}, which are capable of separating different vital signs from the targets. However, ICA requires the dimension of observation to be equal to or higher than the number of sources, which is not always available for radars. Beam sensing-based methods are investigated in mmVital\cite{mmvital}. In this work, reflection loss is calculated to localize and direct the antennas to point toward the target. Similarly, Vimo\cite{vimo} leverages 2-D antenna scanning to localize the object and applies smoothing spline and dynamic programming (DP) to get RR and HR. Its disadvantage is that scanning the beam over the whole space will cost extra time, which will adversely affect real-time performance. Beamforming of MIMO radar has also been found to be effective to enhance the signal from the targets\cite{c33, c34, c35}, but these works do not provide experiments to validate the efficiency in the circumstances of moving interference sources.

	Recently, many data-driven deep learning models\cite{c26, c27} are proposed to filter out the original heartbeat and breath waveform. RF-SCG \cite{ha2020contactless} applies a convolutional neural network (CNN) to reconstruct accurate seismocardiography and effectively overcome the impact of the dynamic environment. MoVi-Fi\cite{chen2021movi} and MoRe-Fi\cite{zheng2021more} respectively exploit multiple layer Multilayer perceptron (MLP) and Variational Encoder-decoder network (VAE) to eliminate the body part motion interference and refine heart and breath waveform. However, these methods always require a large amount of well-labeled training data collected in the RF fields, which is a very time-consuming process.

	\section{Signal Models in Dynamic Environment}
	\subsection{Ideal Signal Models in COTS MmWave Radar Systems}
	The operation of a mmWave radar has been investigated in detail in previous works \cite{c35, c37}. An FMCW radar periodically transmits a series of linearly-increasing frequency signals, called chirps. The period of transmission is $T$, called pulse repetition interval (PRI). Each chirp lasts the time duration $T_c$. During each chirp, the frequency starts from $f_c$ and increases by the bandwidth $B$. Then the transmitted signal is given by 
	\begin{equation}
		x_T (t) = {A_T} e^{j 2\pi (f_c t + \frac{B}{2 T_c} t^2 + \varphi_0)}.  
	\end{equation}
	where $A_T$ is the transmitted amplitude and $\varphi_0$ is the initial phase which is a constant. For a static target located at radial range $R$, after the round-trip delay $\tau = \frac{2R}{c}$ ($c$ is speed of light), the received signal due to the reflected factor $\beta$ is described as
	\begin{equation}
		y (t) = {\beta A_T} e^{j 2\pi (f_c (t - \tau) + \frac{B}{2 T_c} (t - \tau)^2 + \varphi_0)}.
	\end{equation}

	After I/Q demodulation and range correlation, we can omit the constant phase value $\varphi_0$ and  $\frac{B}{2 T_c} \tau^2$ (small in practice) and then obtain
	\begin{equation}
		y(t) = {A_R} e^{j 2 \pi (\alpha R t + \frac{2 R}{\lambda})},
	\end{equation}
	where $\alpha =\frac{2B}{c T_c} $, $\lambda$ is the wavelength of the transmitted signal, $A_{k}$ is the amplitude of response of the received signal.

	In the case of MIMO mmWave radar with a uniform linear array (ULA), the interval between neighboring antennas is $d$. If the target is located at the angle of $\theta$ with respect to the radar, the signal received by the $k^{th}$ antenna is expressed as:
	\begin{equation}
		{y_k(t)} = {A_{k}} e^{j 2 \pi (\alpha R t + \frac{2 R}{\lambda} + \frac{ k d sin\theta}{\lambda})}.	
	\end{equation}

	\subsection{Our Model for Vital Sign Monitoring in Dynamic Scenario}
	\begin{figure}
		\centering
		\includegraphics[width=0.4\textwidth]{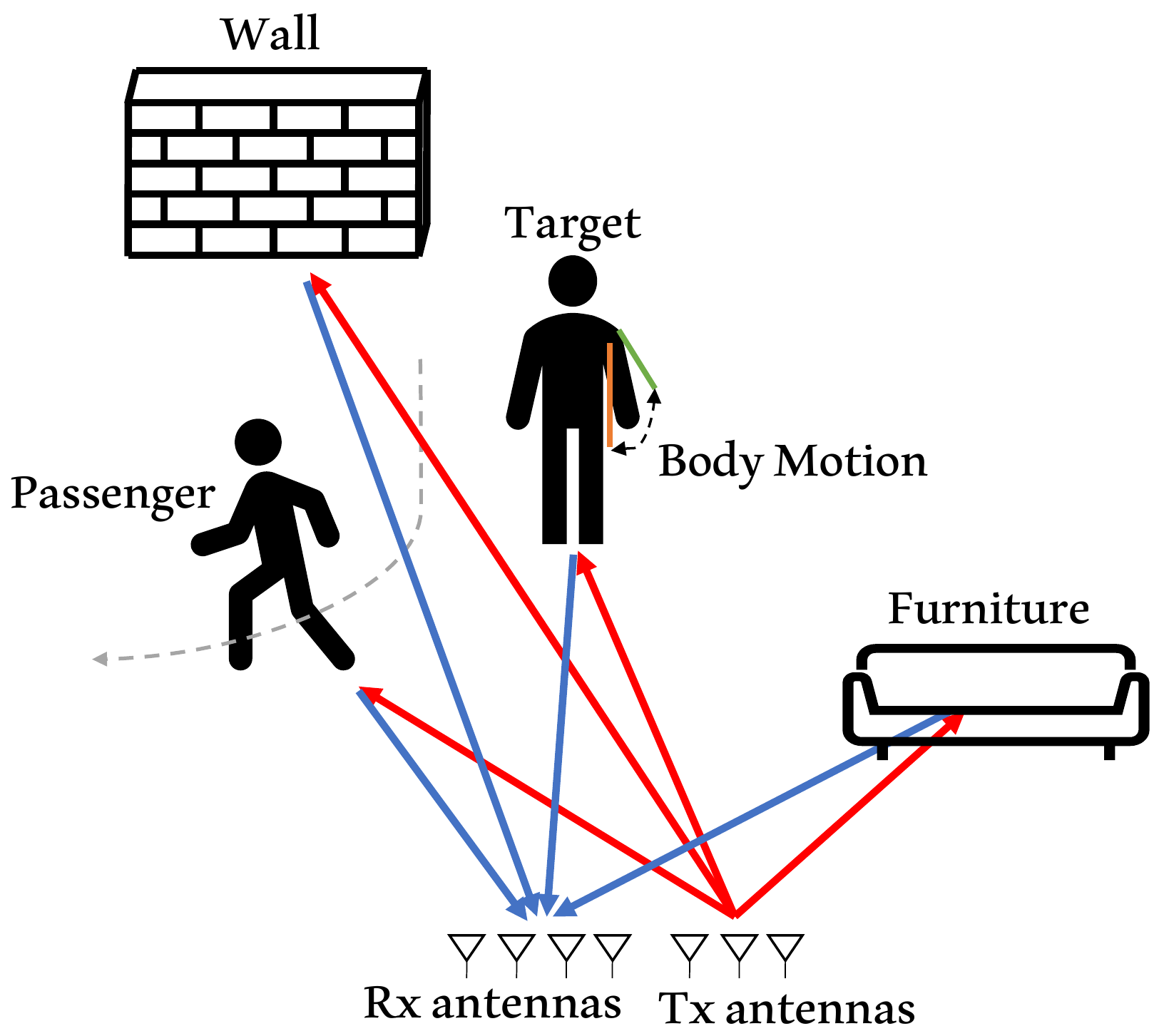}
		\centering
		\caption{Signal propagation in practical environment}
		\label{fig:1}
	\end{figure}

	As shown in Fig.~\ref{fig:1}, the above-mentioned model cannot work well in practice due to the static multipath, the body motion of the target and the dynamic environment (when another person passes by). In this case, we rebuild the signal model by adding the following three interference factors: static objects, target of monitoring and other moving people.
	
	\textbf{1) Static Objects.}
	Assuming that there are $L$ static objects in space and the l-th is located at the range $R_l$ and angle $\theta_l$. The total received signal received by the $k^{th}$ antenna is expressed as:
	\begin{equation}
		{y_{st}(t, k)} =\sum_{l=0}^{L-1} {A_{l, k}} e^{j 2 \pi (\alpha R_l t + \frac{2 R_l}{\lambda} +  \frac{ k d sin\theta}{\lambda})},	
	\end{equation}
	where $A_{l, k}$ is the amplitude of the signal reflected by the $l^{th}$ object and received by the $k^{th}$ antenna.

	\textbf{2) Monitored Target.}
	The signal model that includes the target is more complicated.  
	Since the chest movement and body motion will cause small range variation, even though the target is at the same position, the signal propagation distance $R$ will vary over time. Therefore, we re-write the range as 
	\begin{equation}
		R(t) = R_{tar}(t) + \triangle R_{tar}(t),
	\end{equation}
	where $\triangle R_{tar}(t)$ represents the small time-varying shift caused by the body movement. Here $R_{tar}(t)$ denotes the nominal range at $t$. For a monitored target who stays in the same spot, the nominal range is constant. Thus, the received signal of the target person at the $k^{th}$ antenna can be represented as:
	
	\begin{equation}
		y_{tar}(t, k) =A_{tar, k} e^{j 2 \pi (\alpha R_{tar} t + \frac{2 R_{tar} + 2 \triangle R_{tar}(t)}{\lambda} + \frac{k d sin\theta_{tar}}{\lambda})}.
	\end{equation}
	
	\textbf{3) Other Moving People.}
	For other people walking in space, the signal models not only contain the body motion $\triangle R_{p}(t)$, with a similar expression to that for the target, but also includes the changes on nominal range $R_p(t)$, angle $\theta_p(t)$ and reflected factor. Assuming that there are $P$ people (in addition to the target person) moving in space, the received signal is expressed by:
	\begin{equation}
		y_{m}(t, k) =\sum_{p=0}^{P-1}A_{p, k}(t) e^{j 2 \pi (\alpha R_{p}(t) t + \frac{2 R_p(t) + 2 \triangle R_{p}(t)}{\lambda} + \frac{k d sin\theta_p(t)}{\lambda})}.
	\end{equation}
	
	\textbf{4) Signal Model after Sampling.}
	The signal is sampled and discretized by an Analog/Digital Converter(ADC) after I/Q demodulation. Hence, let 
	\begin{equation}
		t = n T_f + m T_c
	\end{equation}
	be the time of the $n^{th}$ sample in the $m^{th}$ chirp and $T_f$ denotes the sample interval during a chirp. Then the signal can be represented by the fast time $n$, slow time $m$ and antenna $k$ in three dimensions. After simplification, the signal in the dynamic scenario is finally expressed by:
	\begin{equation}
		y_{dy}(n, m, k) = y_{st}(n, m, k) + y_{tar}(n, m, k) + y_m(n, m, k),
	\end{equation}
	where
	\begin{equation}
		\left\{ 
		\begin{aligned}
			&y_{st}(n, m, k) = \sum_{l=0}^{L-1} {A_{l, k}} e^{j 2 \pi (\alpha R_l n + \frac{2 R_l}{\lambda} + \frac{k d sin\theta_l}{\lambda})},\\
			&y_{tar}(n, m, k) \\
			&= A_{tar, k} e^{j 2 \pi (\alpha R_{tar} n + \frac{2 R_{tar} + 2 \triangle R_{tar}(m)}{\lambda} + \frac{k d sin\theta_{tar}}{\lambda})},\\
			\\
			&y_m(n, m, k) \\
			&= \sum_{p=0}^{P-1} A_{p, k}(m) e^{j 2 \pi (\alpha R_{p}(m) n + \frac{2 R_p(m) + 2 \triangle R_{p}(m)}{\lambda} + \frac{k d sin\theta_p(m)}{\lambda})}.  \\
		\end{aligned} 
		\right.
		\label{e1}
	\end{equation}

	
	%
	
	Our model reveals that in order to obtain vital signs, we need to first estimate the target's location (range $R_{tar}$ and angle $\theta_{tar}$) of the target and then extract the phase ($\triangle R_{tar}$). Specifically, moving people and static objects are major hurdles for locating the target. Their signal amplitudes could be larger than the target's and be time-varying (moving people). Also, the moving people will cause noise superimposed on the phase of the target when they walk close to the target.
	
	Furthermore, $\triangle R_{tar}$ contains four major components: (a) heartbeat, (b) respiratory, (c) vibration caused by body part movement, (d) noise caused by moving people nearby. It is noticed that the four components of the received signal could overlap in the frequency domain, which reveals that a classical filter cannot separate vital signs from other components. In the following, we aim to propose a novel scheme to achieve accurate localization and adaptive refinement of vital signs.

	\section{System Overview}
	\begin{figure*}
		\centering
		\includegraphics[width=0.8\textwidth]{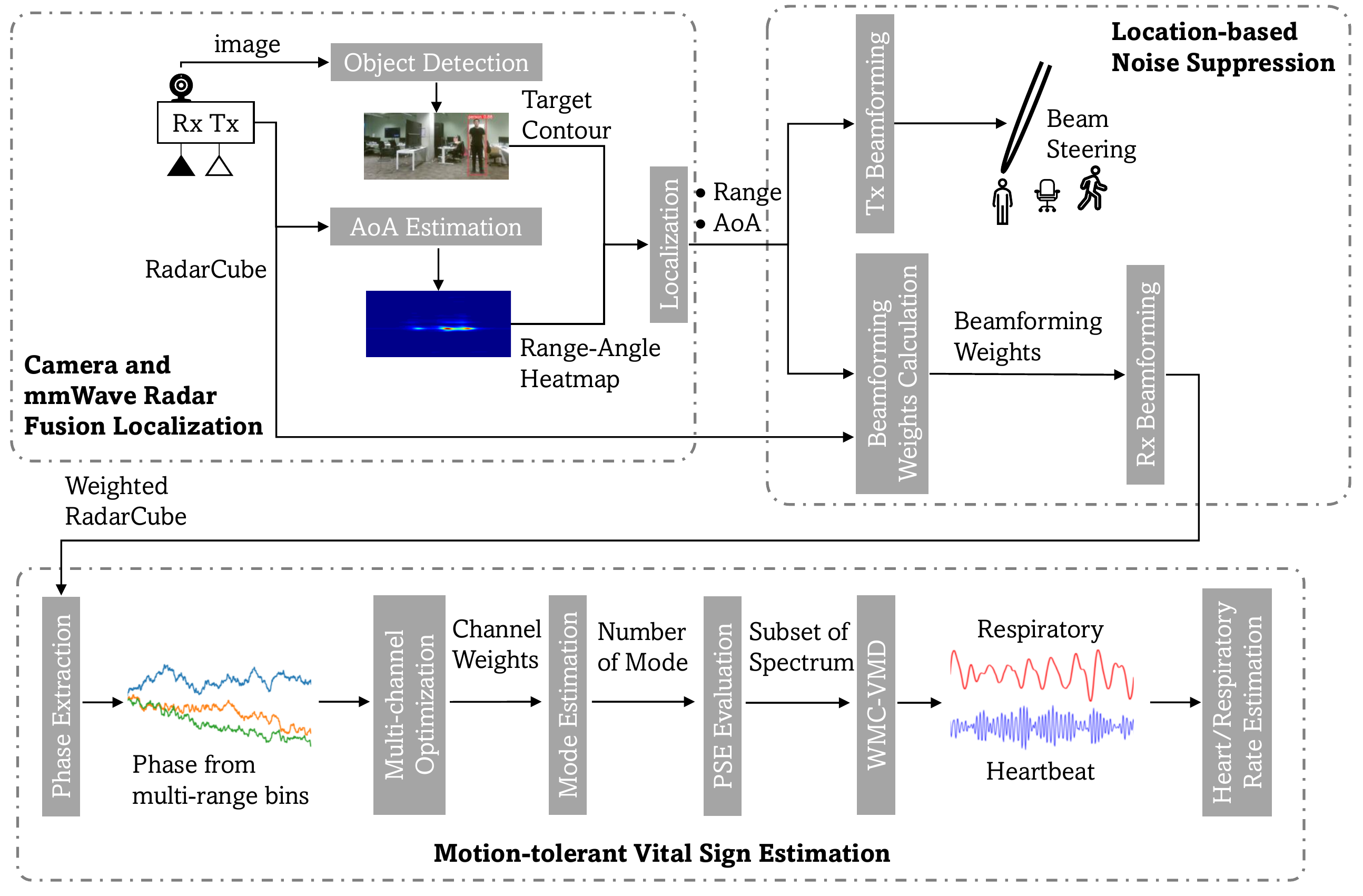}
		\centering
		\caption{System workflow}
		\centering
		\label{fig:3}
	\end{figure*}
	
	This work presents a fusion system \textit{VaR-VSM} that enables dynamic vital sign monitoring using a camera and a MIMO mmWave radar. Our scheme is developed by referring to the COTS mmWave device TI AWR1843. It contains 3 transmitting antennas and 4 receiving antennas. A camera (Kinect V2) is mounted closely with the radar, which works with the radar synchronously. Although Kinect can provide depth + RGB, but only RGB is used in our scheme. There is no difference between Kinect and a conventional RGB camera when Kinect only outputs RGB images. Fig.~\ref{fig:3} illustrates the workflow of our system. The function of each major processing  module is summarized below and will be discussed in detail later.
	
	
	\textbf{1) Camera and mmWave Radar Fusion Localization.} The COTS mmWave radar performs the 1D range FFT on the raw RadarCube in the fast time dimension. This step is directly completed in the radar hardware layer. We use the 1D FFT data output for AoA estimation and obtain the range-angle heatmap. Meanwhile, the camera captures the image stream and we perform a deep-learning based object detection framework on the images. Thus, we fuse the range-angle heatmap and contours of detected persons for accurate localization. The details will be described in Section 4.
	
	\textbf{2) Location-based Noise Suppression.} Once the position of each target is identified, beamforming techniques are adopted to improve the SNR of the signal reflected by the target and mitigate the interference from other people. This consists of Tx-BF and Rx-BF. \textit{Tx-BF} is implemented by shifting the phase of each transmitter. The main beam formed by Tx-BF can point towards the target. \textit{Rx-BF} determines a set of weights that form a receive beam pointing to target too. It enables the received signal to yield maximum powerin this direction. This part will be presented in detail in Section 5. 
	
	\textbf{3) Motion-tolerant Vital Sign Estimation.} To reduce the impact of motion interference, we propose a vital signal reconstruction method by optimizing the variance of the signal phase over successive range bins. Further, we propose the approaches of signal mode estimation and PSE (Power Spectral Entropy) evaluation to automatically measure the mode number of the signal and accelerate the computation. Based on this, heart rate and respiratory rate are calculated by performing FFT over the reconstructed vital sign waveform. The details will be presented in Section 6.

	\section{Camera and mmWave Radar Fusion Localization}	
	Radar-Camera fusion has been investigated in the field of autonomous driving, for tracking or obtaining the high-resolution position of targets \cite{radar_camera1, radar_camera2}. It mainly exploits the high angular resolution of camera and high depth resolution of radar to locate the targets of interest. Inspired by this, we introduce camera to radar sensing to separate location-fixed targets and moving interferers, by leveraging the object detection frame in the angular domain. Different from these works which focus on fusing radar points and camera, fusing range-angle heatmap and camera is proposed in this work. Even when two persons are close to each other, camera can still separate them by the higher angular resolution, which is hardly achievable by mmWave radar with a limited number of antennas. With the assistance of camera, the radar can perform further noise suppression and lower-interference sensing. However, unlike in existing radar-camera fusion, we face new challenges of (1) how to select the static human objects from all the detected objects including moving individuals and static furniture and (2) how to obtain the depth (distance) of the target of interest.
	This section describes a method to pinpoint the subjects of interest by fusing the target contour from images and the range-angle heatmap from the mmWave data.
	
	\subsection{Target Contour Acquisition from Camera}
	\begin{figure}
		\centering
		\includegraphics[width=0.25\textwidth]{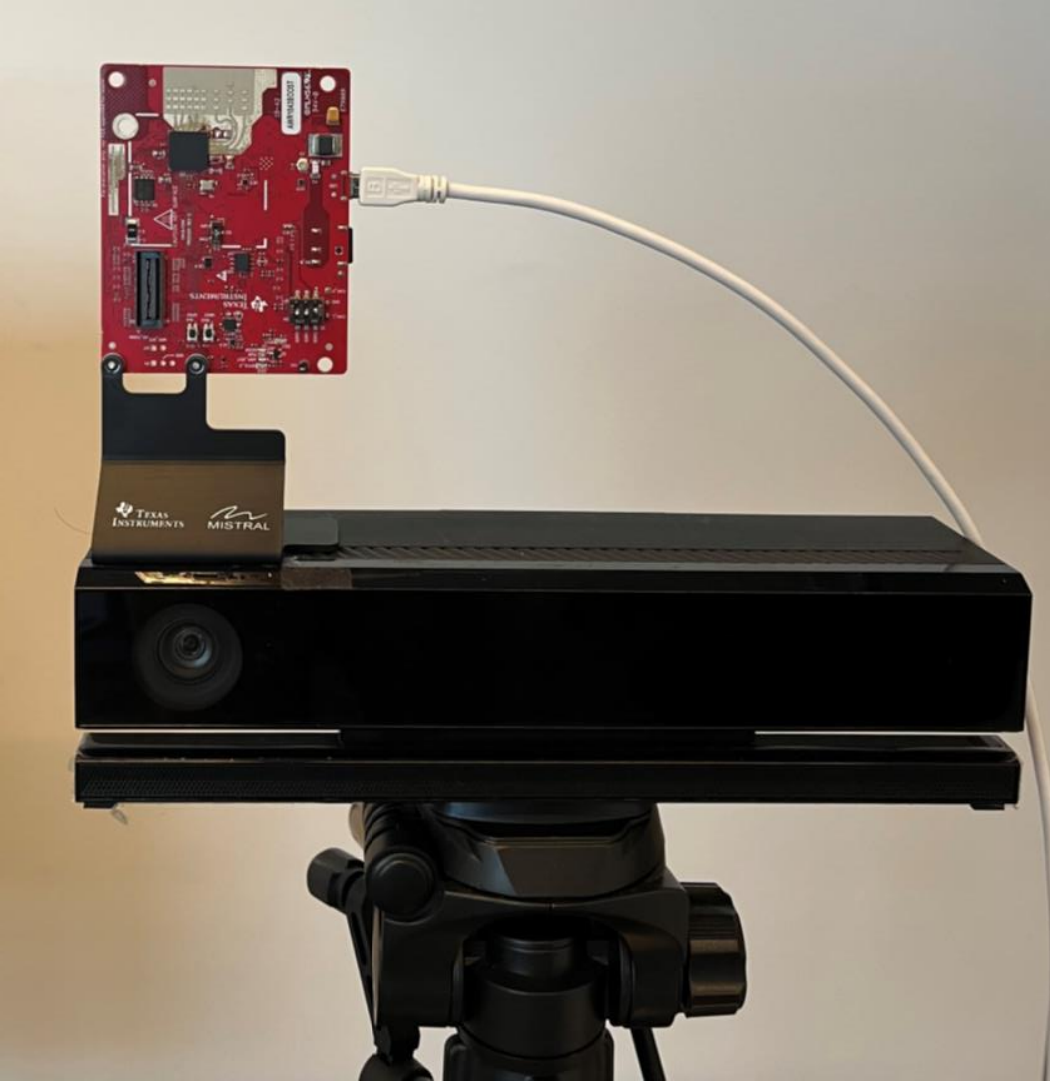}
		\centering
		\caption{Mmwave radar and camera}
		\centering
		\label{figdevice}
	\end{figure}
	The image stream is captured by a camera mounted on the radar (shown in Fig. \ref{figdevice}). The resolution of the image is 1920 $\times$ 1080 pixels. The frame rate is set to 30 FPS. We manually calibrate the orientation and height of the camera lens to align the Field of View (FOV) of the camera with that of the mmWave radar. Once an image is collected, we apply an object detector YOLO V5\cite{c38} to detect all targets of interest, each of which is bounded with a rectangle box on the image. Since the YOLO detector enables to classify as many as 80 classes, there exist many redundant and undesirable bounding boxes containing non-human objects. Hence, those bounding boxes only including the person are automatically selected by reprogramming YOLO. Further, there still exist many similar bounding boxes around each person target. To eliminate the ambiguity of the target of interest, we utilize the non-maximum suppression (NMS)\cite{c38} algorithm to remove those redundant boxes and obtain the optimal one. 
	
	\subsection{Range-angle Heatmap from MmWave Radar}
	\begin{figure}
		\centering
		\includegraphics[width=0.5\textwidth]{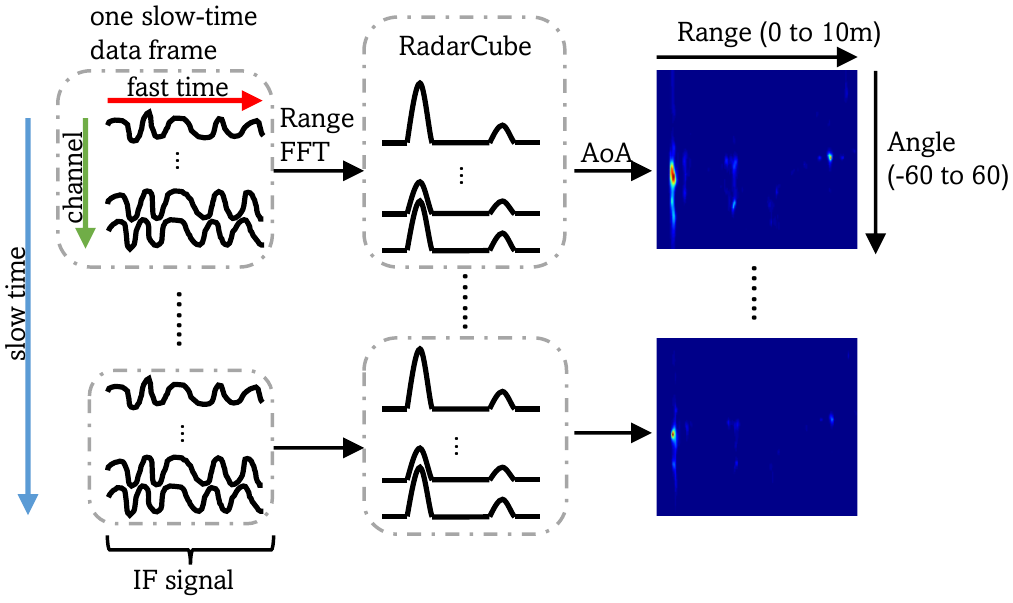}
		\centering
		\caption{Heatmap generation flow}
		\centering
		\label{fig:4}
	\end{figure}

	As shown in Fig.~\ref{fig:4}, the heatmap is generated by stacking the AoA spectrum over all range bins. Firstly, the radar receives the reflected signal and performs I/Q demodulation. The output of this step is called intermediate-frequency (IF) signal. As described in Section 2, the spectrum of range can be estimated by performing FFT on IF signal of all channels over the fast time dimension. In this paper, the RadarCube is defined as the output of this step. The spatial FFT method is commonly used to obtain the AoA spectrum. However, there only exists a small number of receiving elements in the COTS mmWave radar, evenwhen the virtual-array technique is used. For example, the mmWave radar used in our system has a total of 8 virtual receiving antennas in the azimuth direction. The FFT method could not provide a high enough AoA resolution. Although the method of zero-padding is often utilized in FFT, there exists serious spectral leakage and is incapable of separating those closely located objects. Instead, we introduce the minimum variance distortionless response (MVDR)\cite{c39} algorithm to obtain fine-grained AoA spectrum.
	
	\subsection{Stationary Person Localization}
	Once the range-angle heatmap is acquired, the 2D position of the monitored target could be estimated by fusing the contour and peaks in the heatmap. At first, the angular field of view (AFOV) of the camera ranges from -60 degrees to +60 degrees, which is set by the manufacturer. The AFOV of the mmWave radar can cover the range of -90 degrees to +90 degrees. However, the main lobe of the radar transmitter used in the experiment only focuses within the range of about $\pm$60 degrees. The beam gain in other directions is much less than the main lobe's, which makes the radar hard to accurately sense the objects. In this case, the searching range of AoA in the MVDR algorithm is set to $\pm$60 degrees, thereby achieving the AFOV matching between the camera and mmWave radar.
	
	\begin{figure}
		\centering
		\includegraphics[width=0.5\textwidth]{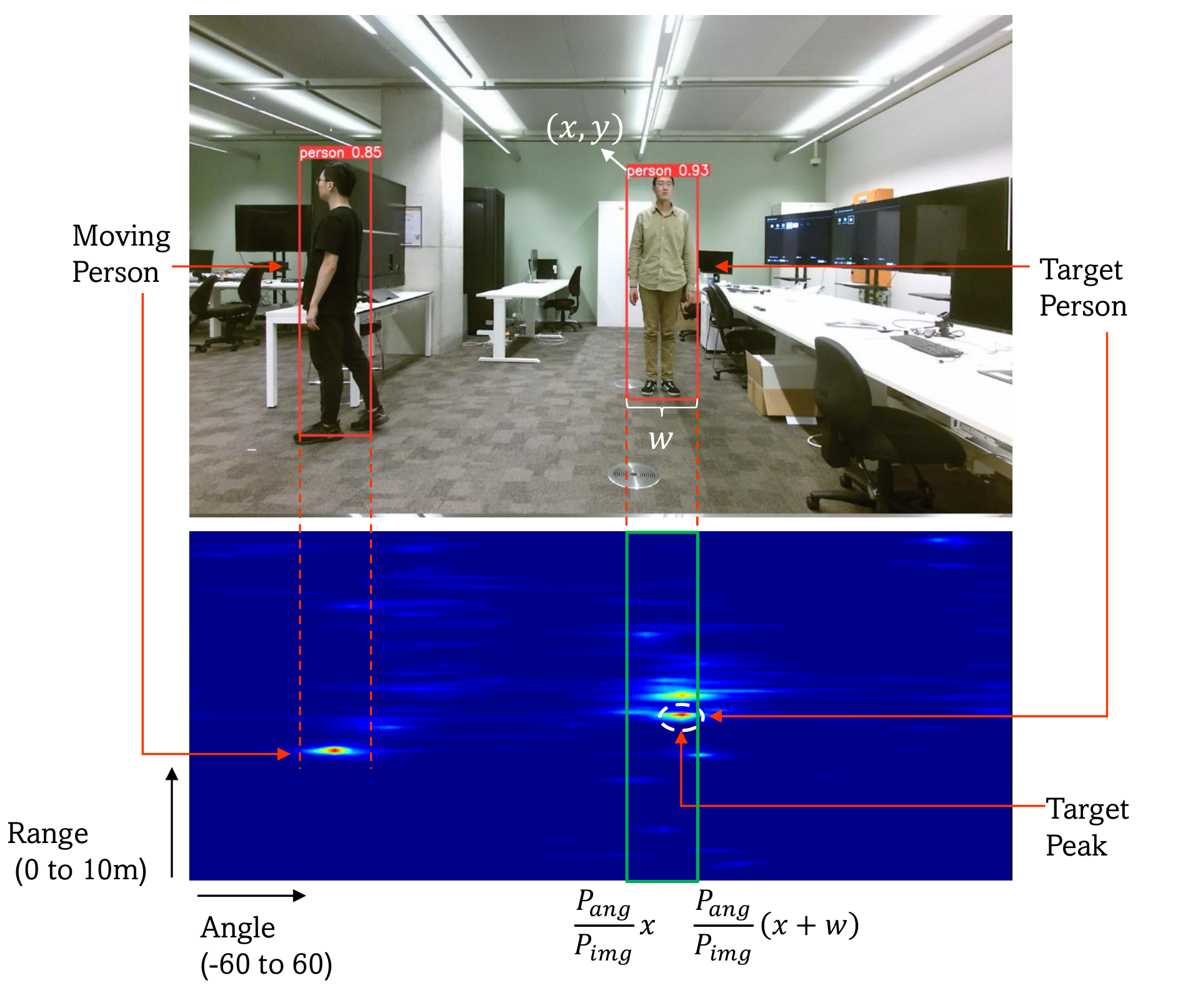}
		\caption{Localization}
		\label{fig:5}
		\vspace{-1.5em}
	\end{figure}
	
	\begin{figure*}
		\centering
		\begin{subfigure}{0.245\textwidth}
			\centering
			\includegraphics[width=0.7\textwidth]{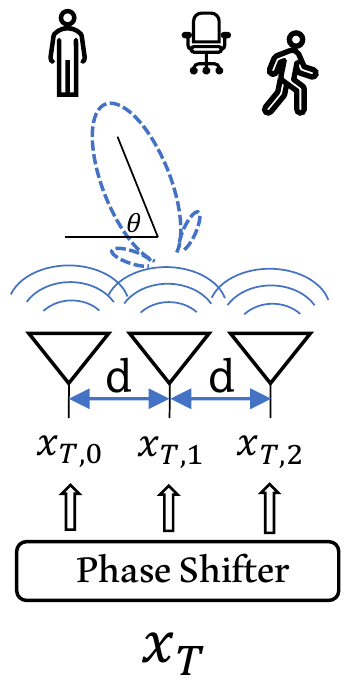}
			\subcaption{BF Illustration}
			\label{fig:6a}
		\end{subfigure}
		\begin{subfigure}{0.245\textwidth}
			\centering
			\includegraphics[width=0.9\textwidth]{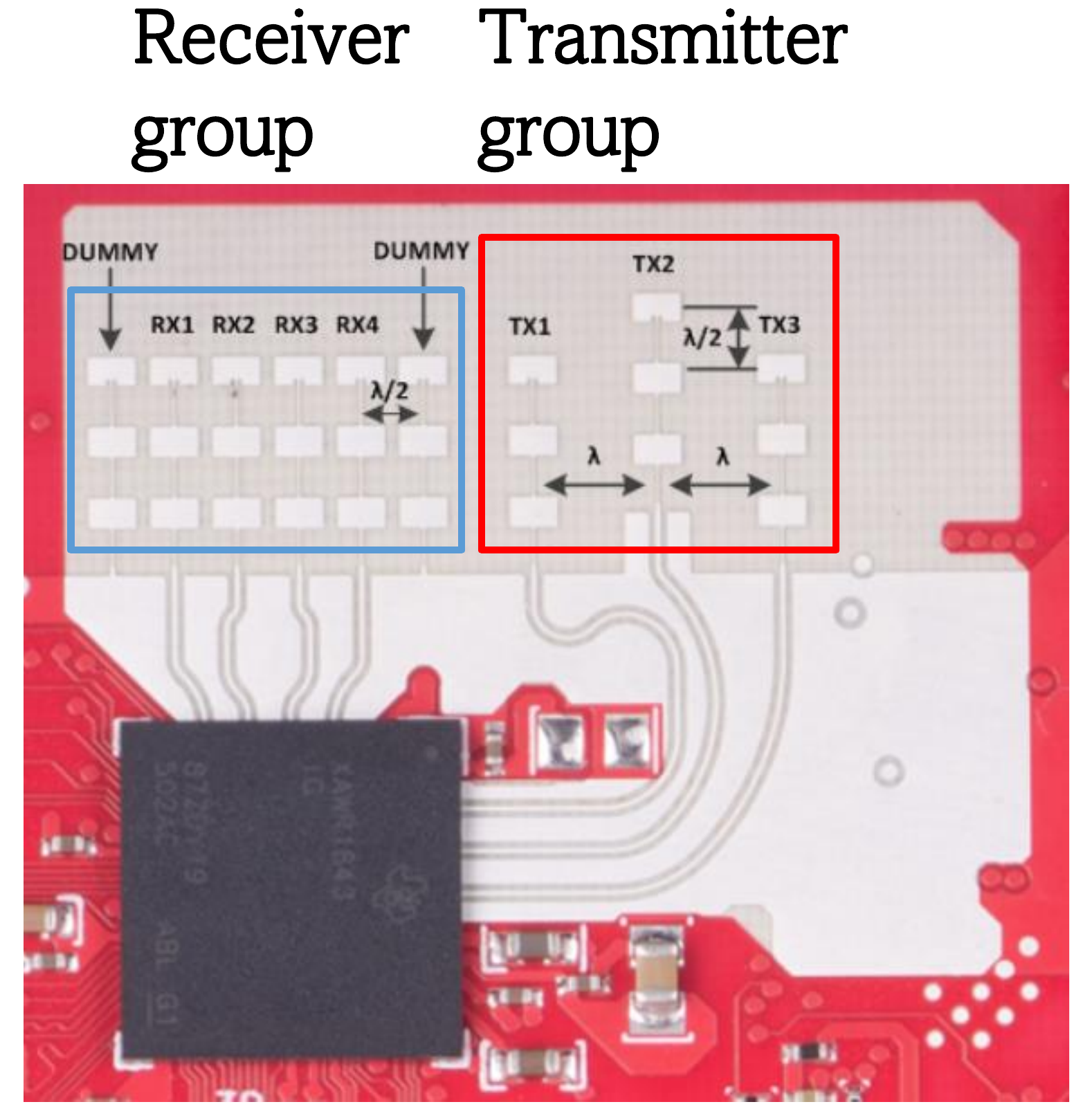}
			\subcaption{Antenna geometry}
			\label{fig:6b}
		\end{subfigure}
		\begin{subfigure}{0.245\textwidth}
			\centering
			\includegraphics[width=1.025\textwidth]{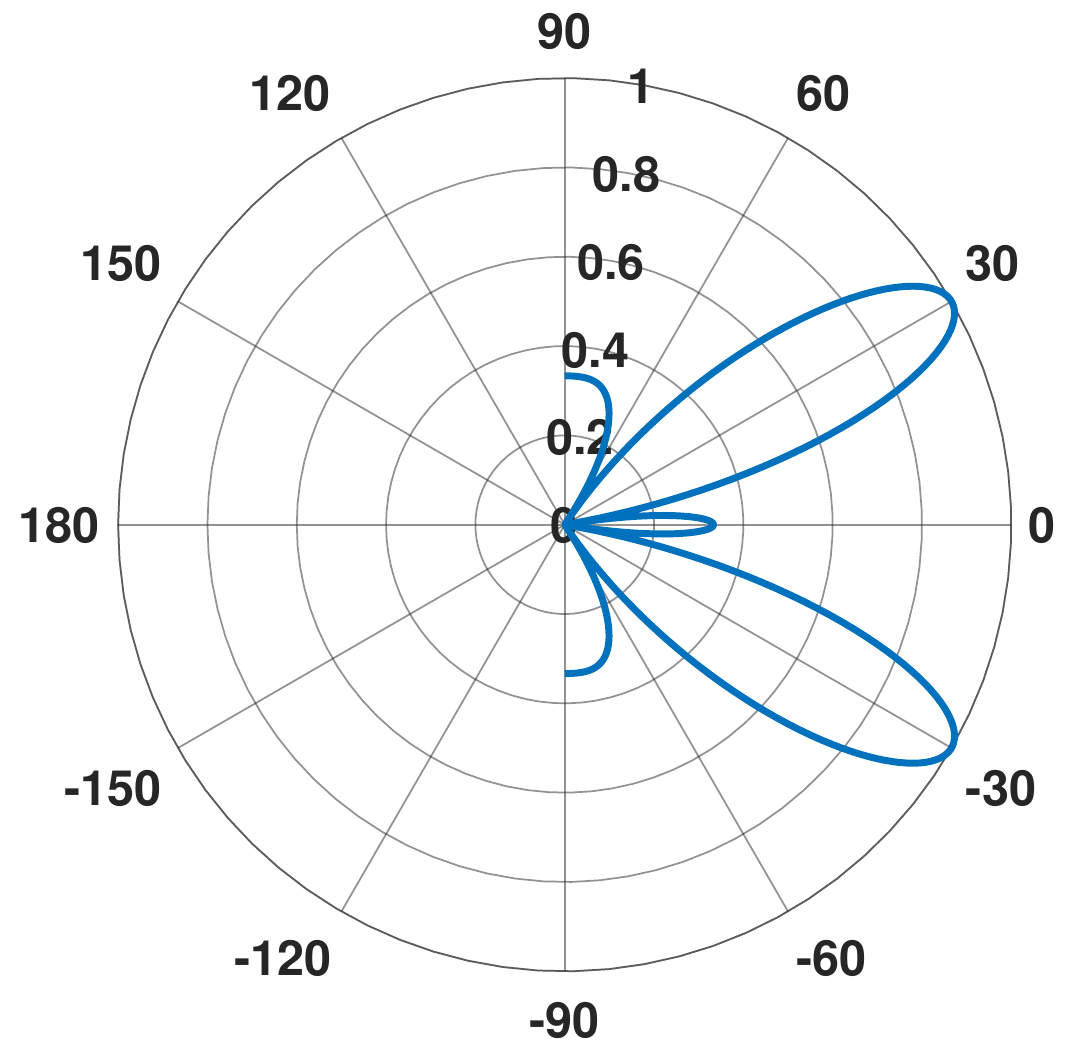}
			\subcaption{Tx-BF of $\lambda$ spaced elements}
			\label{fig:6c}
		\end{subfigure}
		\begin{subfigure}{0.245\linewidth}
			\centering
			\includegraphics[width=1.025\textwidth]{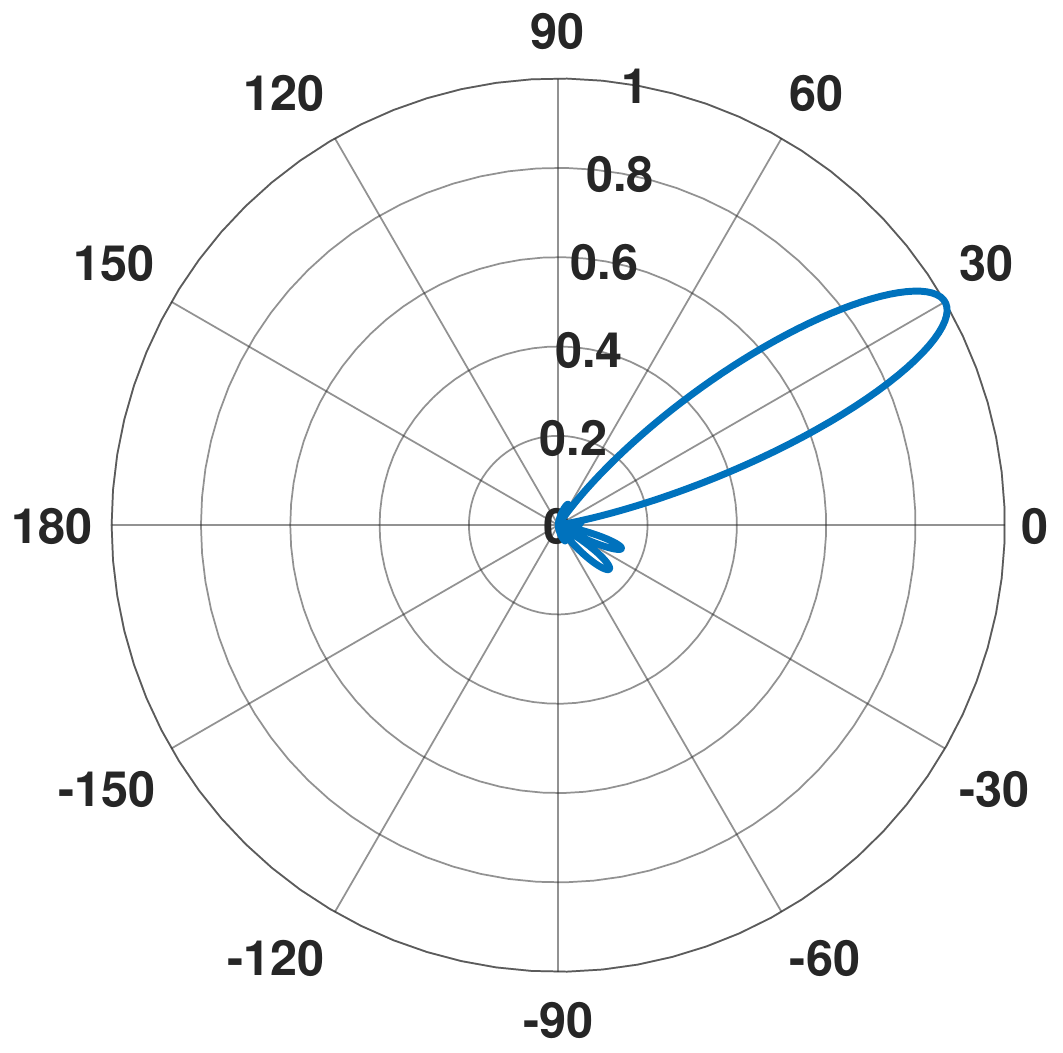}
			\subcaption{Rx-BF}
			\label{fig:6d}
		\end{subfigure}
		\vspace{-1em}
		\caption{Beamforming in AWR1843}
		\label{fig:6}
		\vspace{-1.5em}
	\end{figure*}
	
	In real environments, there may exist multiple persons. We leverage the camera data to filter out the stationary targets of interest for vital sign recognition. Note that only the locations of these targets are required to be unchanged, while their body parts may still have some movement. The output of the image-based YOLO detector includes the coordinate of the start point $(x, y)$, width $w$ and height $h$ of all people contours. When a person is moving, its start point and width may change over time. For the target person these values are almost constant. In this case, we can set two thresholds to remove those bounding boxes of moving targets. If the change in the start points (or widths) of the bounding boxes corresponding to a person is larger than the predefined threshold, we consider him/her as the moving target and thereby ignore the bounding box. After that, we can map the pixels of the stationary persons in the image into the angle bins on the heatmap. The optimal AoA of each target can be obtained by searching the maximum values within the width of the corresponding bounding box. This can be expressed by:
	
	\begin{equation}
		R_{tar}, \theta_{tar} = \underset{{\theta}\in \frac{P_{ang}}{P_{img}} \left[ x, x + w \right]}{\underset{R \in \left[0, 10 \right]} {\mathop{\text{argmax}}}} \bm{H_m}(R, \theta), 
	\end{equation}
	where $\bm{H_m}$ denotes the range-angle heatmap estimated by MVDR (0 to 10 represents the corresponding range values on the heatmap and $\theta$ is obtained by remapping the image pixels to the heatmap). $P_{ang}$ and $P_{img}$ respectively represent the numbers of angle bins of range-angle heatmap and pixels of the image.

	\section{Location-based Noise Suppression}
	In this section, we exploit the location-based beamforming technique to suppress the impact of range bin overlapping caused by moving persons. They may have the same distance from a static person to the mmWave radar but in different directions. Tx-BF and Rx-BF are combined to handle this issue.
	
	\subsection{Beamforming at Transmitter}
	Tx-BF can physically control the transmitter beam to form the desired beam pattern. The main beam lobe can be changed to steer towards the direction of interest by controlling the phase of the transmitting signal on each antenna (shown in Fig.~\ref{fig:6a}). 
	
	The COTS mmWave radar used in our system has three Tx antennas spaced by $d$. As described above, the position of each stationary target to be tracked is calculated. To steer the beam towards the target direction $\theta_{tar}$, we firstly define the phase-shift beamforming weights vector,
	\begin{equation}
		\bm{w}_{tx} = [1,  e^{j 2\pi \frac{d}{\lambda} sin\theta_{tar}}, e^{j 2\pi \frac{2d}{\lambda} sin\theta_{tar}}].
	\end{equation}
	Then we multiply the original transmitting signal $x_T (t)$ by the beamforming weight $\bm{w}_{tx}$ to obtain
	\begin{equation}
		\left\{
		\begin{aligned}
			x_{T,0} (t) &= {A_T} e^{j 2\pi (f_c t + \frac{B}{2 T_c} t^2 + \varphi_0)}  \\
			x_{T,1} (t) &= {A_T} e^{j 2\pi (f_c t + \frac{B}{2 T_c} t^2 + \frac{d}{\lambda} sin\theta_{tar} + \varphi_0)} \\
			x_{T,2} (t) &= {A_T} e^{j 2\pi (f_c t + \frac{B}{2 T_c} t^2 + \frac{2d}{\lambda} sin\theta_{tar} +\varphi_0)}
		\end{aligned} 
		\right..
	\end{equation}
	In this case, the radar can transmit the main beam in a specific direction $\theta_{tar}$. It can enhance the signal strength of the stationary person in a certain direction and attenuate the reflected signals of moving persons in other directions.
	
	Unfortunately, if the spacing between two adjacent antennas is larger than half a wavelength, the modified radiation pattern has multiple main beams. Fig.~\ref{fig:6b} illustrates the antenna layout of the radar in our experiment. The distance between each transmitter is $\lambda$, the wavelength of the carrier signal. The radar generates two main lobes, shown in Fig.~\ref{fig:6c}. The desired direction is 30 degrees while there is an extra beam pointing towards -30 degrees. In this case, if there is an undesired object located at -30 degrees, the signal will still be interfered since there is no attenuation on the beam response from that angle. In the following, we will use the receiving beamforming in the digital domain to further suppress the interference from other undesired directions.
	
	\subsection{Beamforming at Receiver}
	Rx-BF is performed over the receiving antennas. It aims to use a set of weights to maximize the power from the specific direction in digital domain. 
	
	To maximize the received signal power for each target, we can use the estimated AoA $\theta$ to compute the beamforming weights $\bm{w}_{rx}$, i.e., 
	\begin{equation}
		\bm{w}_{rx} = [1,  e^{j 2\pi \frac{d_r}{\lambda} sin\theta}, ..., e^{j 2\pi \frac{(L - 1)d_r}{\lambda} sin\theta}],
	\end{equation}
	where $L$ and $d_r$ denote the number and distance of receiving antennas, respectively. 
	
	Finally, we use the above RX weights to combine all received signals and then obtain
	\begin{equation}
		\bar{y}(t) = \sum_{l=0}^{L-1} w_{rx}^{*}(l)x_l(t) = \bm{w}_{rx}^{H} \bm{x}(t),
	\end{equation}
	where the symbols $*$ and $H$ denote the conjugate and conjugate transpose operations, respectively. Fig.~\ref{fig:6d} shows the extra lobe is eliminated by Rx-BF and the overall beam concentrates the direction of 30 degree.

	\section{Motion-tolerant Vital Sign Estimation}
	This section describes the proposed weighted multi-channel variational mode decomposition (WMC-VMD) and uses it to estimate vital signs. The optimization weights calculation, parameter selection and acceleration methods are also introduced in this section. 
	
	\subsection{Vital Sign Estimation by Weighted Multi-channel Variational Mode Decomposition}
	To estimate vital signs, we need to extract phase from the target's range bin by phase demodulation and compensation\cite{c33, c40}. However, there are at least 4 components contained in extracted phase sequence: heartbeat, respiratory, body movement and range overlapping. Although the target has been separated by beamforming, there are still two issues.
	
	Firstly, the short-term body motion of the target may cause interference signals with frequencies close to those of the respiratory and heartbeat signals. Therefore, traditional filters, such as Butterworth filter, cannot accurately extract vital signs. The VMD algorithm \cite{c36} has the potential to address this issue. It aims to estimate the Intrinsic Mode Functions (IMFs) with the assumption that each mode of the signal is around a central frequency with a narrow bandwidth. Then It leverages Alternating Direction Method of Multipliers (ADMM) to concurrently calculate the central frequency and mode function. 
	
	However, another issue is that the slight body motion may result in the change of the predefined range bin calculated in Section 5. It is hard to predict which range bin the person will be located at in each sampling time. The only prior knowledge is that the nominal position of the target is roughly unchanged. Therefore, it is necessary to fuse the phase sequence from the adjacent range bins of the target. In this case, the VMD becomes ineffective since it only supports the decomposition of the single-dimension signal. The improved MS-VMD\cite{c20} is capable of dealing with the high-dimension signals by equally adding up each dimension in the optimization procedure. But this operation cannot pinpoint the signal from the true range bin the target is located in.
	
	In this work, we propose weighted multi-channel VMD (WMC-VMD), which is capable of adaptively combining the phase sequence from multiple range bins to estimate the IMFs. In WMC-VMD, we introduce a set of adaptive weights $w_l$ (the notation $l$ denotes the $l^{th}$ range bin) to sum up the phase sequence from different range bins. These weights can eliminate the offsets among different range bins and remap the phase sequence in the same scale. Since the phase sequence from each range bin contains the vital sign, each sequence is optimized individually. 
	
	The proposed algorithm WMC-VMD will be detailed in Algorithm 1,  consisting of the following steps: 1) applying Hilbert transform in phase sequence to obtain the analytic signal, 2) for each mode $u_{k}$, mixing it with a complex exponential of the center frequency to shift its spectrum to the baseband, 3) different with VMD-based methods, the baseband of $u_{k}$ is estimated by squared $l^2$-norm of the gradient with the operation of weighted summation on each phase sequence from multi-range bins. Specifically, WMC-VMD is modeled as: 
	\begin{equation}
		\begin{gathered}	
			\underset{{{u_k}, {w_k}}}{\text{min}} \sum_{k=1}^{K} \sum_{l=1}^{L} {\left \| \partial_t [(\delta(t) + \frac{j}{\pi t}) * {u_k}(t)] e^{-j \omega_k t} \right\|}_2^2 \\ 
			s.t. \sum_{k=1}^{K} u_k(t) = {w_l}{{s_l}(t)},
		\end{gathered}
	\end{equation}
	where $K$ and $L$ respectively represent the number of modes and channels, $w_l$ denotes the adaptive weight (the calculation will be introduced in Section 7.2), and $s_l(t)$ represents the phase sequence of the $l^{th}$ range bin. The augmented Lagrangian function is formed to solve this problem. With the constraints of weighted multiple linear equations, it becomes:
	\begin{equation} \label{Eq18}
		\begin{aligned}
			&\mathscr{L} = \sum_{k=1}^{K} \sum_{l=1}^{L} {\left \| \partial_t [(\delta(t) + \frac{j}{\pi t}) * {u_k}(t)] e^{-j \omega_k t} \right\|}_2^2 \\
			& + \sum_{l=1}^{L} {\left \| {w_l}{{s_l}(t)} - \sum_{k=1}^{K} u_k(t) \right\|}_2^2
			+ \sum_{l=1}^{L} \langle {\lambda_l (t)}, {{w_l}{s_l}(t) - \sum_{k=1}^{K} u_k(t)} \rangle.
		\end{aligned}
	\end{equation}
	
	This task is divided into updating the IMF and center frequency of each mode. Based on Parseval/Plancherel Fourier isometry under the $l^2$-norm, we can convert this optimization problem from the time domain to frequency domain and perform iterative sub-optimization by ADMM, the final updating results of IMF and center frequency can be obtained as
	\begin{equation} \label{Eq19}
		{{\hat{u}_k}^{n+1}}(\omega) =
		\frac{{\sum^{L}_{l=1}}{[{w_l { S_{l}}}(\omega) - {\begin{matrix} \sum_{i\neq k} {\hat{u}_i}^{n+1} (\omega) \end{matrix}} + \frac{\hat{\lambda}^{n}_{l} (\omega)}{2}]}}{L + 2 (\omega - {\omega_k}^n)^2},
	\end{equation}
	and 
	\begin{equation}
		{{\hat{\omega}_k}^{n+1}} =
		\frac{\begin{matrix} \int_{0}^{\infty} \omega {\left| {\hat{u}_k}^{n+1} (\omega) \right|}^2 \, d\omega\end{matrix}}
		{\begin{matrix} \int_{0}^{\infty}{\left| {\hat{u}_k}^{n+1} (\omega) \right|}^2 \, d\omega\end{matrix}}.
	\end{equation}
	

	\subsection{Weights Calculation}
	In the above section, a set of adaptive weights are introduced to fuse the phase sequences from range bins. Specifically, when a target is staying stationary, the range bins around the target all contain the vital signs but with different amplitudes. If there is a small range shift, other range bins will have stronger vital signs. Since this process could occur at any time, the phase sequence from one range bin could be suppressed for a moment while others are on the contrary. Based on this, we propose a method that aims to identify a set of weights $\bm{w}$ to minimize the variance of signal from multiple channels. By multiplying these weights in WMC-VMD, the variability of phase sequence from all range bins will be normalized to the same scale to improve the decomposition results. To calculate the weights, we define the following optimization problem:
	\begin{equation}
		\begin{gathered}
			\mathop{min}\limits_{\bm{w}} \ {\bm{w}^T}{\bm{s} \bm{s}^T}{\bm{w}} \\
			s.t. \sum_{l=0}^{L-1} {w_l} = 1,
		\end{gathered}
	\end{equation}
	where $\bm{s} \in {R}_{L \times N}$ is $N$ phase samples extracted from $L$ continuous range bins.
	
	To solve this constrained optimization problem, we apply the Lagrange multipliers to obtain the weight $\bm{w}$,
	\begin{equation}
		\bm{w} = \frac{{(\bm{s}\bm{s}^T)^{-1}}\bm{I}}{{\bm{I}^T} (\bm{s}\bm{s}^T)^{-1} \bm{I}},
	\end{equation}
	where $\bm{I}$ is a vector with a size of $1 \times L$ and each element is 1.

	\subsection{Parameter Selection}
	In the existing VMD-based algorithm, the number of modes $K$ needs to be identified as a prior. The selection of $K$ will significantly impact on the decomposition and optimization. Intuitively, $K$ is supposed to be 4 (heartbeat, respiratory, body movement and passenger's range overlapping) based on the empirical assumption for the number of signal sources. However, this cannot work efficiently in this work because in the real environment, when the target moves the body parts, the mode number cannot be guaranteed to remain unchanged. For instance, the target may move the arms together with the legs (even the torso). Since the movements caused by arms and legs are much more salient than heartbeat and breath, $K$ needs to be at least 5 instead of 4. If $K$ were still set to 4, the decomposition results would only contain the IMFs of arms, legs, range overlapping and respiratory but miss heartbeat. 
	
	In WMC-VMD, we leverage singular spectrum analysis (SSA) to automatically select $K$. At first, the Hankel matrix is constructed by reforming the weighted phase sequence signal:
	\begin{equation}
		\bm{X} = 
		\left[
		\begin{matrix}
			s_1 & s_2 & s_3 & \cdots & s_n \\
			s_2 & s_3 & s_4 & \cdots & s_{n+1} \\
			\vdots & \vdots & \vdots & \ddots & \vdots \\
			s_m & s_{m+1} & s_{m+2} & \cdots & s_{N}\\
		\end{matrix}
		\right], 
	\end{equation}
	where $N$ is the total number of samples, $n=N-m+1$ and $M$ is the window length. After that, the singular value decomposition (SVD) is applied on $\bm{H} = \bm{XX}^T$, and we can obtain
	\begin{equation}
		\bm{H} = \bm{U \Sigma V}^T, 
	\end{equation}
	where $\bm{\Sigma}$ is the singular value matrix which is rectangular diagonal and the elements are sorted in the decreasing order. We choose the mode number K as the number of the largest values having 70+\text{percentile} power out of the total singular values.

	\subsection{Computation Acceleration}
	
	\begin{figure}
		\centering
		\begin{minipage}[t]{0.49\linewidth}
			\centering
			\includegraphics[width=1\textwidth]{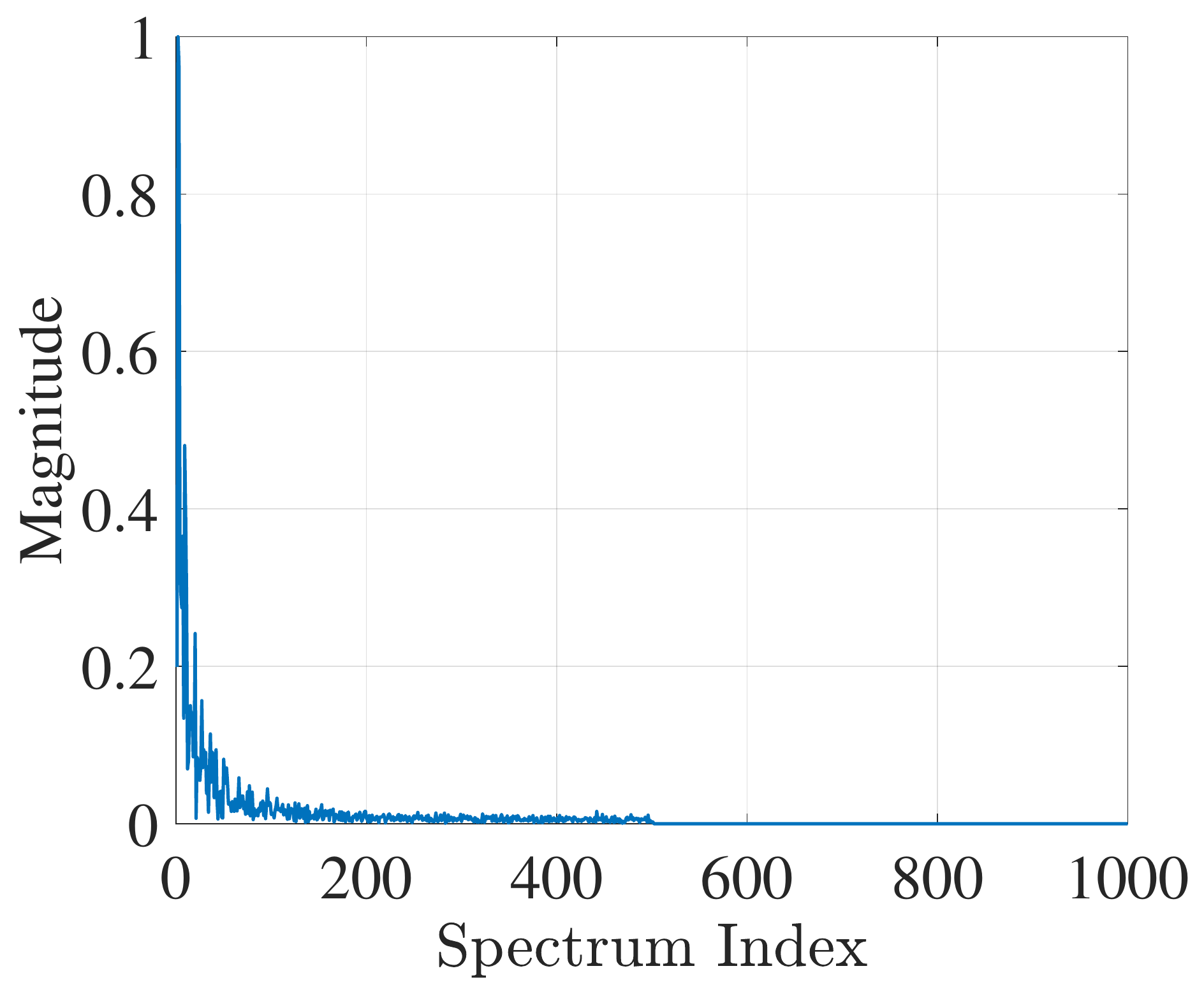}
			\subcaption{Spectrum of extracted phase sequence}
			\label{fig:11a}
		\end{minipage}
		\begin{minipage}[t]{0.49\linewidth}
			\centering
			\includegraphics[width=1\textwidth]{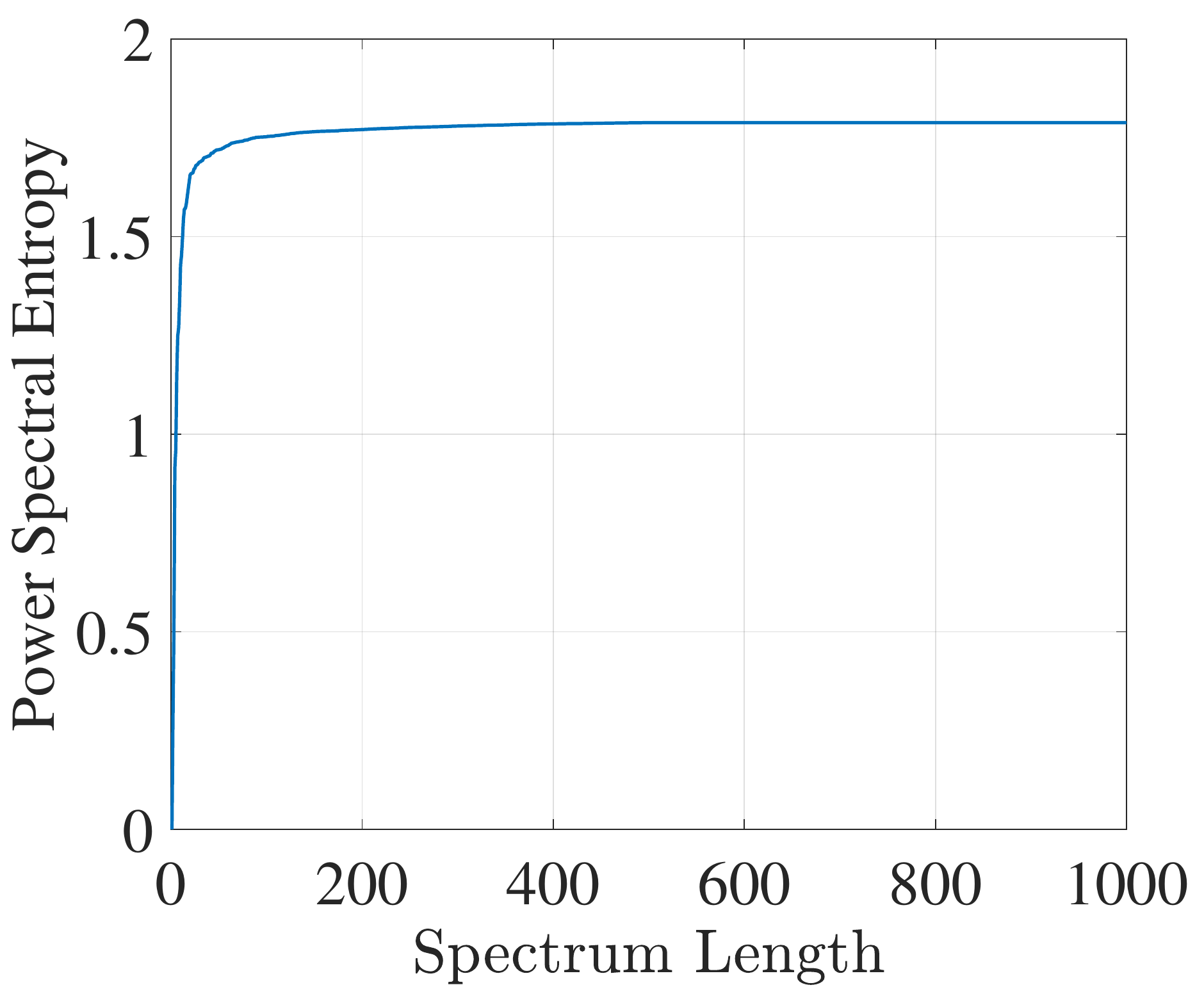}
			\subcaption{Visualization of PSE}
			\label{fig:11b}
		\end{minipage}
		\vspace{-1em}
		\caption{PSE Analysis}
		\label{fig:11}
	\end{figure}
	
	In WMC-VMD, the optimization is performed in the frequency domain, which requires the Fourier Transform operation over the whole snapshots of samples. From Eqs. \ref{Eq18} and \ref{Eq19}, we can observe that in each iteration, the entire spectral components are used to calculate and update the new IMF and central frequency. Although this process can obtain precise decomposition results, it is time-consuming and storage-costly. Fortunately, the spectrum of the extracted phase sequence is very similar to the tail distribution (shown in Fig. \ref{fig:11a}), which implies that the main information is gathered in the low-frequency domain. Therefore, to accelerate the optimization and save computation resources, the power spectral entropy (PSE) is introduced to show that the subset of the original spectrum is capable of reconstructing IMF and estimating central frequency effectively. The PSE of a signal is always utilized to measure the information containing in its spectrum based on Shannon entropy \cite{shannon1948mathematical}. The steps are shown below:
	
	1) Calculate the power spectral density (PSD) of the whole signal and then normalize it:
	\begin{equation}
		P(\omega_i) = \frac{1}{N} {\left| X(\omega_i)\right|}^2
	\end{equation}
	where $N$ is the total number of frequency bins.
	
	2) Normalize the PSD to get the probability density:
	\begin{equation}
		p_i = \frac{P(\omega_i)}{\sum_{i} P(\omega_i)}
	\end{equation}
	
	3) Calculate the PSE:
	\begin{equation}
		PSE = -\sum_{i=1}^{N} p_i ln{p_i}
	\end{equation}
	
	Fig.~\ref{fig:11b} illustrates the PSE of extracted phase sequence. When the size of the subspace is chosen to be 1000, corresponding to the entire spectrum, the PSE reaches the highest. On the other hand, the PSE for size 100 is very close to that for the whole spectrum. This reveals that we can choose the first 100 components, which is only $1/10$ of the whole spectral sequence, to replace the entire spectral components used in (18) and (19), thereby the computation time and resources are dramatically reduced.
	
	\begin{algorithm}
		\caption{Vital Sign Decomposition based on WMC-VMD}
		\label{alg:1}
		\begin{algorithmic}
			\STATE {$\bm{w} = \frac{{(\bm{s}\bm{s}^T)^{-1}}\bm{I}}{{\bm{I}^T} (\bm{s}\bm{s}^T)^{-1} \bm{I}}$}
			\STATE {$\bm{s}_{c}(t) = Hilbert(\bm{s}(t))$}
			\STATE {$\bm{S}(\omega) = FFT(\bm{s}_c(t))$}
			\STATE {${\hat{\bm{S}}}(\omega) \gets choose \ main \ components \ of \ \bm{S}(\omega)$}
			\STATE {$K \gets Singular \ Spectrum \ Analysis$}
			\STATE {Initialize $\left\{ {\hat{u}_i}^1 \right\}$, $\left\{ {\hat{\omega}_i}^1 \right\}$, ${\hat{\lambda}^1}$, $q$ $\gets 0$}
			\WHILE{$n \gets n + 1$}
			\FOR {$k = 1 : K$}
			\STATE {Update $\hat{u}_k$ for all $\omega \geq 0$:}\\
			${{\hat{u}_k}^{n+1}}(\omega) \gets$ \\
			$\frac{{\sum^{L}_{l=1}}{[{w_l \hat{ S_{l}}}(\omega) - {\begin{matrix} \sum_{i<k} {\hat{u}_i}^{n+1} (\omega) \end{matrix}} - {\begin{matrix} \sum_{i>k} {\hat{u}_i}^{n} (\omega) \end{matrix}} + \frac{\hat{\lambda}^{n}_{l} (\omega)}{2}]}}{1 + 2\alpha (\omega - {\omega_k}^n)^2}$
			\STATE {Update $\omega_k$:}\\
			${{\hat{\omega}_k}^{n+1}} \gets \frac{\begin{matrix} \int_{0}^{\infty} \omega {\left| {\hat{u}_k}^{n+1} (\omega) \right|}^2 \, d\omega\end{matrix}}
			{\begin{matrix} \int_{0}^{\infty}{\left| {\hat{u}_k}^{n+1} (\omega) \right|}^2 \, d\omega\end{matrix}}$
			\ENDFOR
			
			\FOR{$l = 1 : L$}
			\STATE {Dual ascent for all $\omega \geq 0$:}\\
			${\hat{\lambda}}^{n+1}_{l} (\omega) \gets {\hat{\lambda}}^{n}_{l} (\omega) + \eta ({w_l \hat{S}_{l}}(\omega) - \sum_{k}{{\hat{u}_i}^{n+1}}(\omega))$
			\ENDFOR
			
			\IF {Convergence: $\frac{{{\left \| {\hat{u}_k}^{n + 1} - {\hat{u}_k}^{n} \right\|}_2}^2}{{{\left \| {\hat{u}_k}^{n} \right\|}_2}^2} < \epsilon$}
			\STATE{break;}
			\ENDIF
			\ENDWHILE
			\STATE{${\Tilde{u}}(\omega) \gets zero \ padding \ {\hat{u}}(\omega) \ to \ original \ length$}
			\STATE {${\Tilde{u}}(t) = IFFT(\Tilde{u}(\omega))$}
			
		\end{algorithmic}
	\end{algorithm}

	\section{Implementation and Evaluation}
	In this section, we will conduct a series of experiments to validate our scheme.
	
	\subsection{Implementation}
	
	\textbf{System Setup.} The device and experimental setup are shown in Fig. \ref{figdevice}. The TI 1843 mmWave radar and the webcam are mounted together. The center of Rx and Tx pairs is aligned at the camera lens vertically. Both of them are fixed on a tripod at a height of 1.3 m. 
	
	\textbf{Experiment Environment.} The system is deployed in our office shown in Fig. \ref{fig:13c}, containing multiple chairs, desks, computers, etc. The environment can induce serious multipath interference. Other testing environments include a meeting room and kitchen, shown in Fig. \ref{rest}.
	
	\textbf{Evaluation.} A target to be tracked sits on a chair or stands. The target is allowed to move the body in a period, which generates motion interference. At the same time, multiple persons walk around the office and induce dynamic interference. The target person is located in a different position in each experiment. Five volunteers are recruited as the target person and moving interference in turn. 
	
	\textbf{Baselines.} We compare our scheme with two state-of-the-art works based on SVD+ICA\cite{c29} and long-short-term-memory (LSTM)\cite{gong2021rf}. Specifically, the work\cite{c29} applies SVD subspace decomposition to improve the localization and ICA to obtain vital signs. And the work\cite{gong2021rf} utilizes LSTM-based deep neural network to estimate heart and breath rates in dynamic scenarios. 
	
	\textbf{Ground Truth.} The ground truth of the respiration rate and heartbeat rate are captured using a sport watch.
	
	\textbf{Metric.} The error is defined as the absolute value of the difference between the estimated heartbeat rate (HR) and respiration rate (RR) $R_e$ and their corresponding ground truth rate $R_g$, i.e., $|R_e - R_g|$. The measurement units for breath and heartbeat rate are the respirations per minute (RPM) and beats per minute (BPM), respectively.
	
	\textbf{Computation Platform.} The system is implemented in Python 3.7 and C and runs on a desktop with Intel(R) Core CPU i7-7700 3.6 GHz A4 and 32G memory.

	\subsection{Overall Performance and Comparison}
	
	\begin{figure}
		\centering
		\begin{minipage}[t]{0.49\linewidth}
			\centering
			\includegraphics[width=1\textwidth]{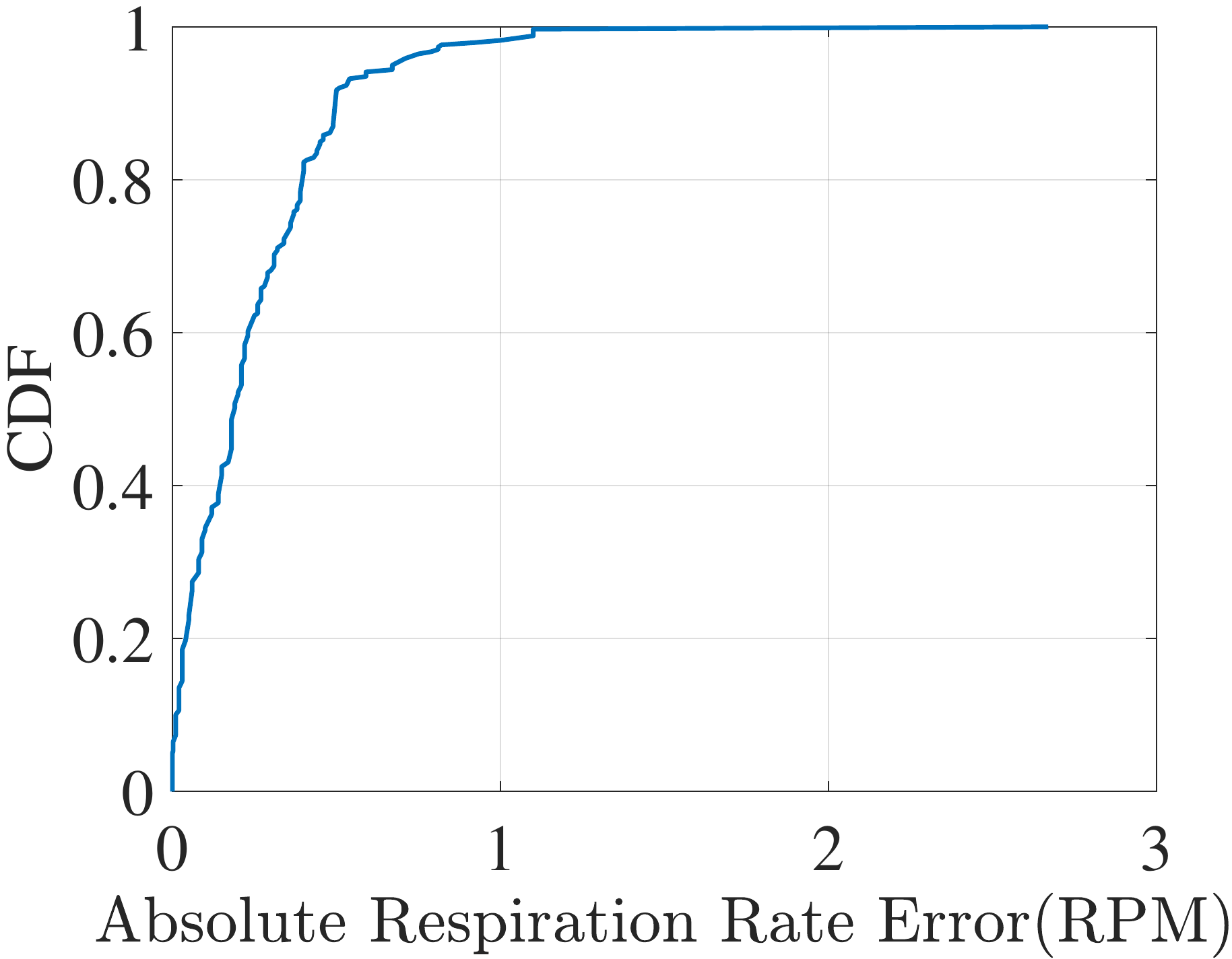}
			\subcaption{Respiration}
			\label{fig:12a}
		\end{minipage}
		\begin{minipage}[t]{0.49\linewidth}
			\centering
			\includegraphics[width=1\textwidth]{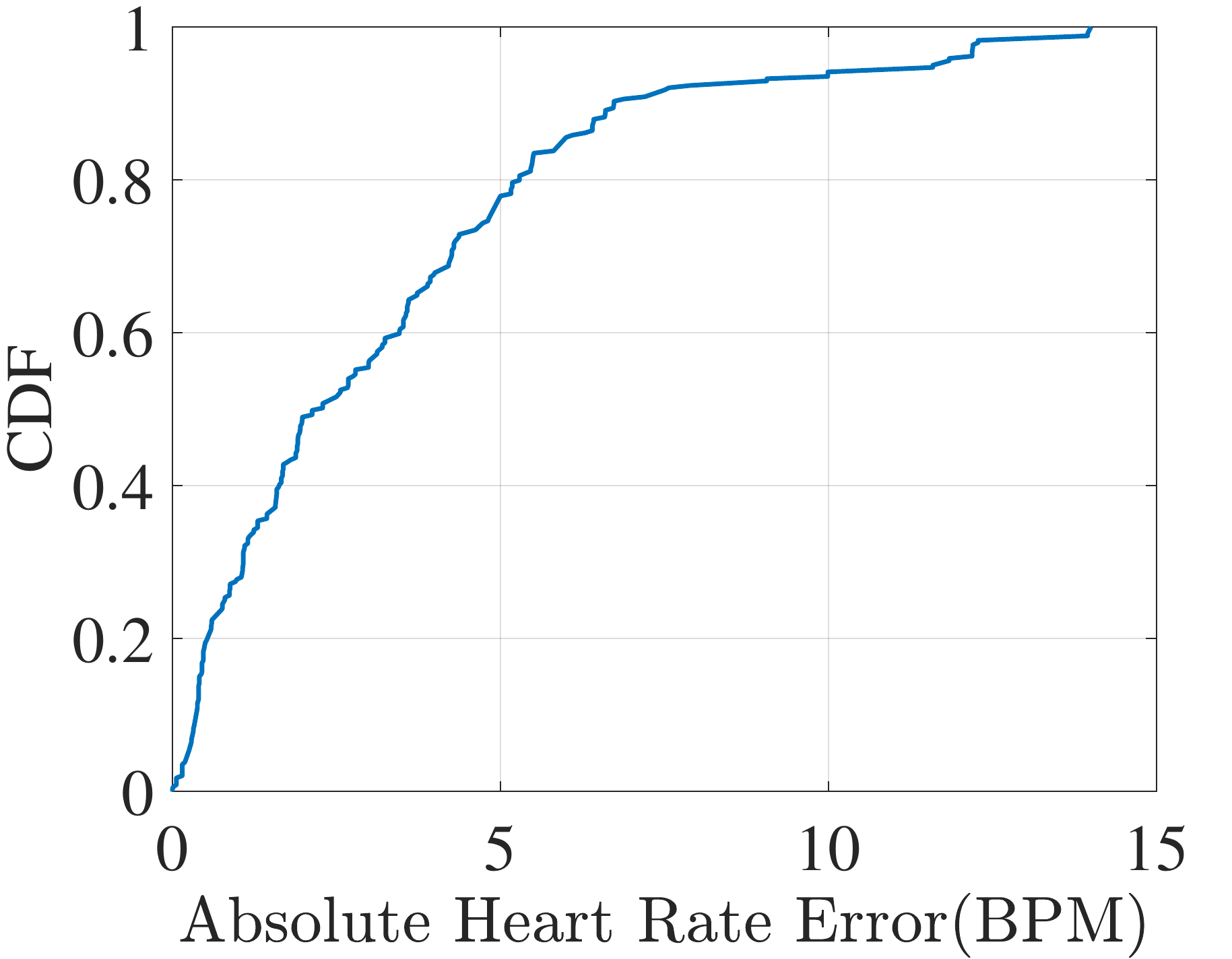}
			\subcaption{Heart}
			\label{fig:12b}
		\end{minipage}
		\vspace{-0.8em}
		\caption{Overall Performance}
		\label{fig:12}
	\end{figure}
	
	\begin{table}
		\caption{Comparison of overall estimation errors}
		\vspace{-1.5em}
		\label{table_cmp}
		\center
		\begin{tabular}{c|c|c}
			\hline
			Methods & RR (RPM)   & HR (BPM)  \\ \hline
			SVD + ICA\cite{c29}     & 1.5  & 3    \\ \hline
			LSTM\cite{gong2021rf}      & 3.22 & 2.86 \\ \hline
			This work & 0.19 & 2.29 \\ \hline
		\end{tabular}
		\vspace{-1em}
	\end{table}
	
	\begin{table}
		\caption{Estimation errors under different distances}
		\vspace{-1.5em}
		\label{table1}
		\center
		\begin{tabular}{c|c|c|c|c}
			\hline
			\multicolumn{1}{c|}{\multirow{2}{*}{Methods}} & \multicolumn{2}{c|}{RR (RPM)} & \multicolumn{2}{c}{HR (BPM)}  \\ \cline{2-5} 
			\multicolumn{1}{c|}{}                       & \multicolumn{1}{c|}{2m} & 3m & \multicolumn{1}{c|}{2m} & 3m \\ \hline
			SVD + ICA\cite{c29}                         & \multicolumn{1}{c|}{1.52}  & 1.55  & \multicolumn{1}{c|}{3.54}  & 3.43  \\ \hline
			LSTM\cite{gong2021rf}                       & \multicolumn{1}{c|}{5.27}  & 4.44  & \multicolumn{1}{c|}{2.93}  & 1.93  \\ \hline
			This work                                   & \multicolumn{1}{c|}{0.22}  & 0.45  & \multicolumn{1}{c|}{1.91}  & 2.49  \\ \hline
		\end{tabular}
		\vspace{-0.6em}
	\end{table}
	
	\begin{table}
		\caption{Estimation errors for different motions}
		\vspace{-1.5em}
		\label{table2}
		\center
		\begin{tabular}{c|c|c|c|c}
			\hline
			\multicolumn{1}{c|}{\multirow{2}{*}{Methods}} & \multicolumn{2}{c|}{RR (RPM)} & \multicolumn{2}{c}{HR (BPM)}  \\ \cline{2-5} 
			\multicolumn{1}{c|}{}                       & \multicolumn{1}{c|}{Periodical} & Random & \multicolumn{1}{c|}{Periodical} & Random \\ \hline
			SVD + ICA\cite{c29}                       & \multicolumn{1}{c|}{$>$5}  & $>$5 & \multicolumn{1}{c|}{-}  & - \\ \hline
			LSTM\cite{gong2021rf}                       & \multicolumn{1}{c|}{3.56}  & 2.17  & \multicolumn{1}{c|}{7.36}  & 4.62  \\ \hline
			This work                                   & \multicolumn{1}{c|}{1.44}  & 1.17  & \multicolumn{1}{c|}{6.61}  & 5.5  \\ \hline
		\end{tabular}
		\vspace{-1em}
	\end{table}
	We test our system by conducting a series of experiments under different ranges between the target and radar, body motion types, number of moving disruptors, and different environments (i.e., office, meeting room and kitchen). We use an empirical Cumulative Distribution Function (CDF) of the absolute error to evaluate the overall system performance, shown in Fig.9. We can see that 90 percent of respiration experiments yield less than 0.5 RPM error. And 80 percent of heartbeat experiments can achieve less than 5 BPM errors. The median errors in respiration rate (RR) and heartbeat rate (HR) estimation are 0.19 RPM and 2.29 BPM, respectively. 
	
	Next, we compare our work with two state-of-the-art works based on SVD+ICA\cite{c29} and deep learning based LSTM\cite{gong2021rf}. In fairness to these two works, the comparison is under the same range and body motion type. Table \ref{table_cmp} indicates the overall error rates of the three works. It can be observed that our work achieves the lowest RR error rate, which is nearly 8 times less than that of SVD+ICA and 17 times less than that of the LSTM based works, respectively. Also, the HR error rate of our work is 25 percent and 31 percent less than those of SVD+ICA and LSTM based works.
	
	Furthermore, we conduct two fine-grained comparison experiments.
	
	(1) We conduct comparisons at the same distances between our work and SVD+ICA\cite{c29} and LSTM\cite{gong2021rf}. According to Table \ref{table1}, our method achieves over 7 times and 3 times improvements on RR errors at 2 and 3 meters respectively, compared with SVD + ICA. For HR our method outperforms SVD+ICA by approximately 1.9 and 1.4 times. Our method achieves comparable performance with deep learning based LSTM (about 0.5 BPM higher in 3m/HR but 30 percent less in 2m/HR and 10 times less in RR). It should be pointed out that the maximum working range of our work is about 7m, as will be detailed in Section 8.7, and the SVD+ICA fails when the range is over 3.5m.

	(2) We further compare the performance using the same body motions with those for SVD+ICA\cite{c29} and LSTM\cite{gong2021rf}. As shown in Table \ref{table2}, Our method still achieves nearly half of the error rates in RR estimation for both motion types compared with LSTM. In terms of HR, even though our method does not perform better than LSTM for random motion, it still achieves a 10 percent lower error rate for periodical movement. The observation indicates that LSTM can be more effective for random noise suppression since the deep learning based algorithm is more efficient in learning and fitting specific patterns from pseudo random observations. In contrast, our proposed VMD based methods perform better in estimating patterns from periodical signals. It is noted that LSTM requires a large amount of data for prior training while our method does not.
	
	In addition, SVD+ICA fails in HR sensing when the target has body motion even if the distance between target and radar is very small. This indicates that ICA based methods are more fragile when the signal contains interference from different (body parts) sources due to the limited observation dimensions. In contrast, our proposed WMC-VMD is more effective in handling such interference.

	\subsection{Impact of Beamforming}
	\begin{figure}
		\centering
		\begin{minipage}[t]{0.75\linewidth}
			\centering
			\includegraphics[width=1\textwidth]{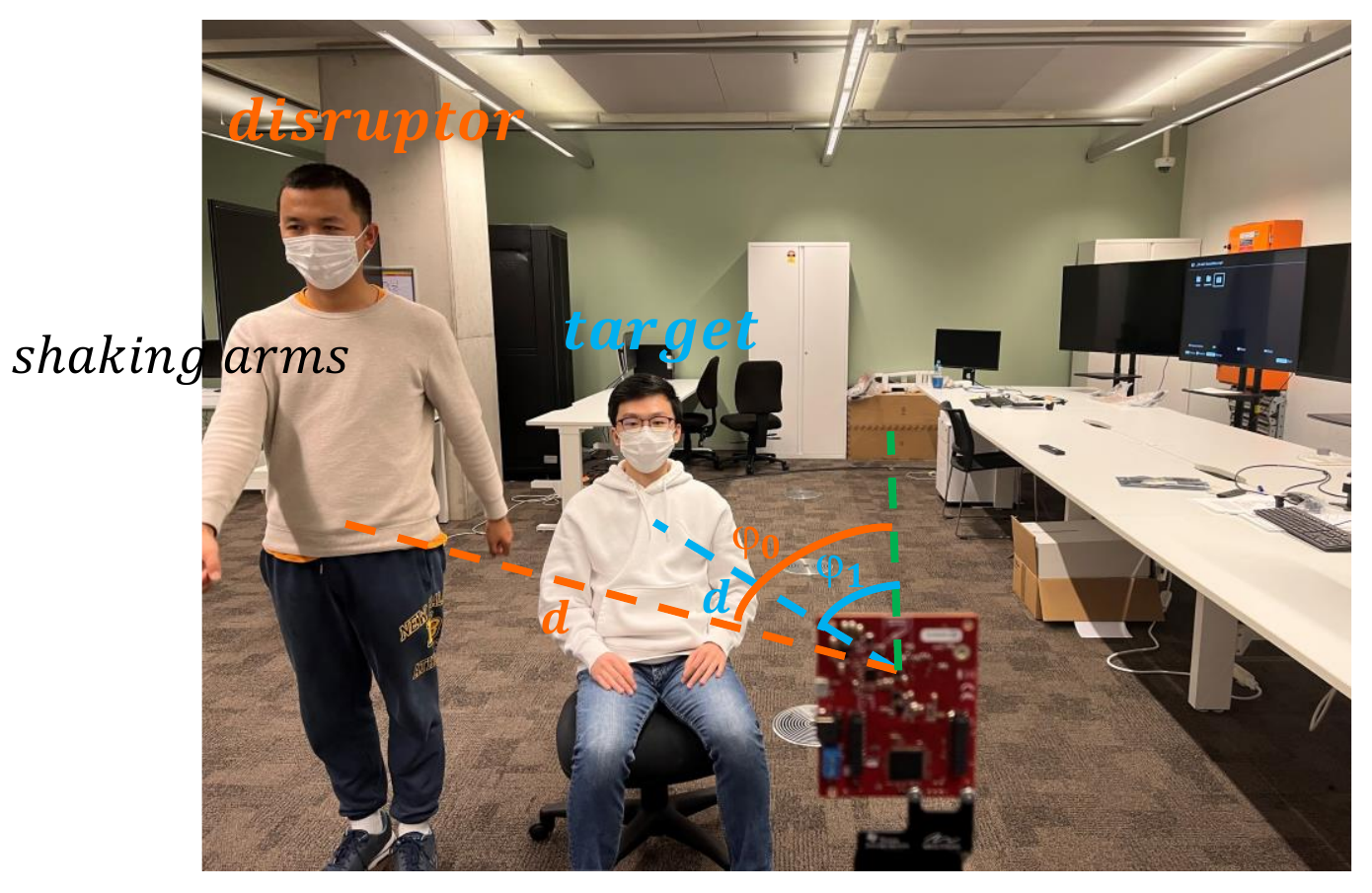}
			\subcaption{Beamforming Validation Scene}
			\label{fig9a}
		\end{minipage}
		\begin{minipage}[t]{1\linewidth}
			\centering
			\includegraphics[width=0.8\textwidth]{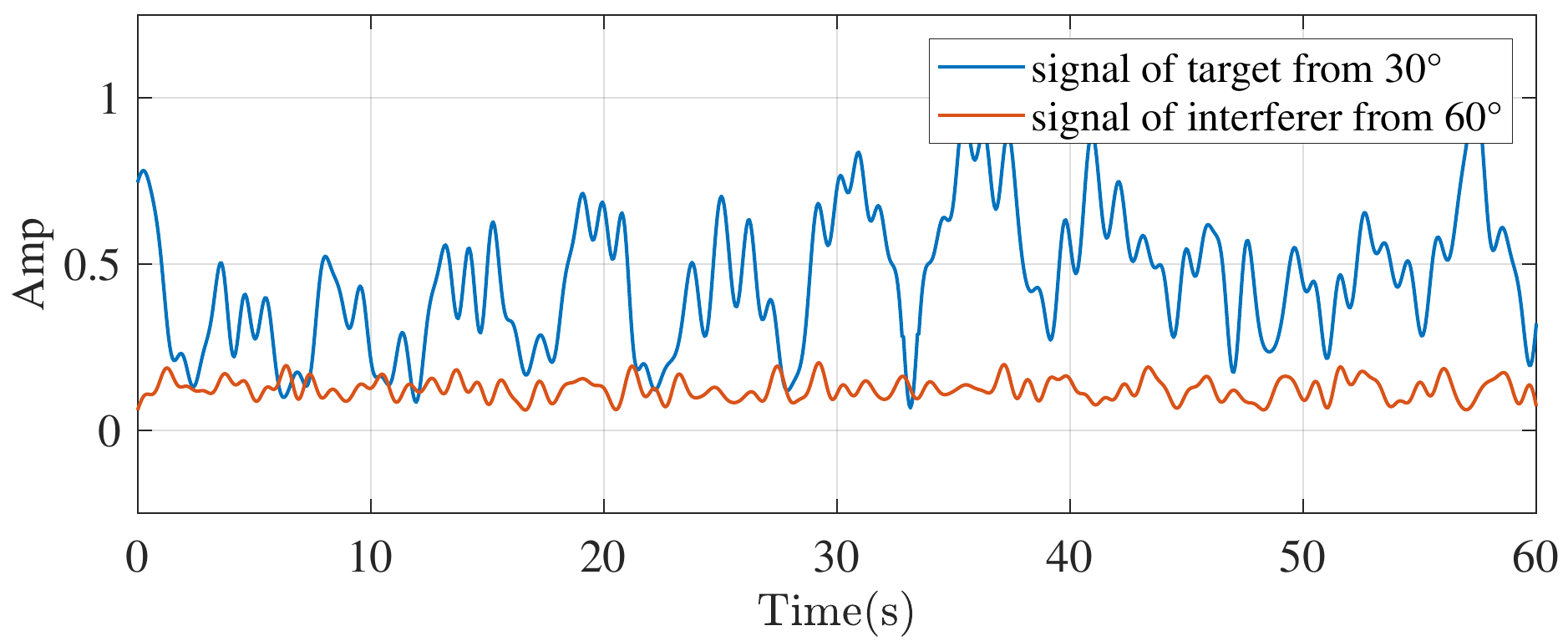}
			\subcaption{Received Signal after Beamforming}
			\label{fig9b}
		\end{minipage}
		\caption{An Example of Beamforming}
		\label{fig:9}
	\end{figure}
    \begin{figure}
    	\centering
    	\begin{minipage}[t]{0.49\linewidth}
    		\centering
    		\includegraphics[width=1\textwidth]{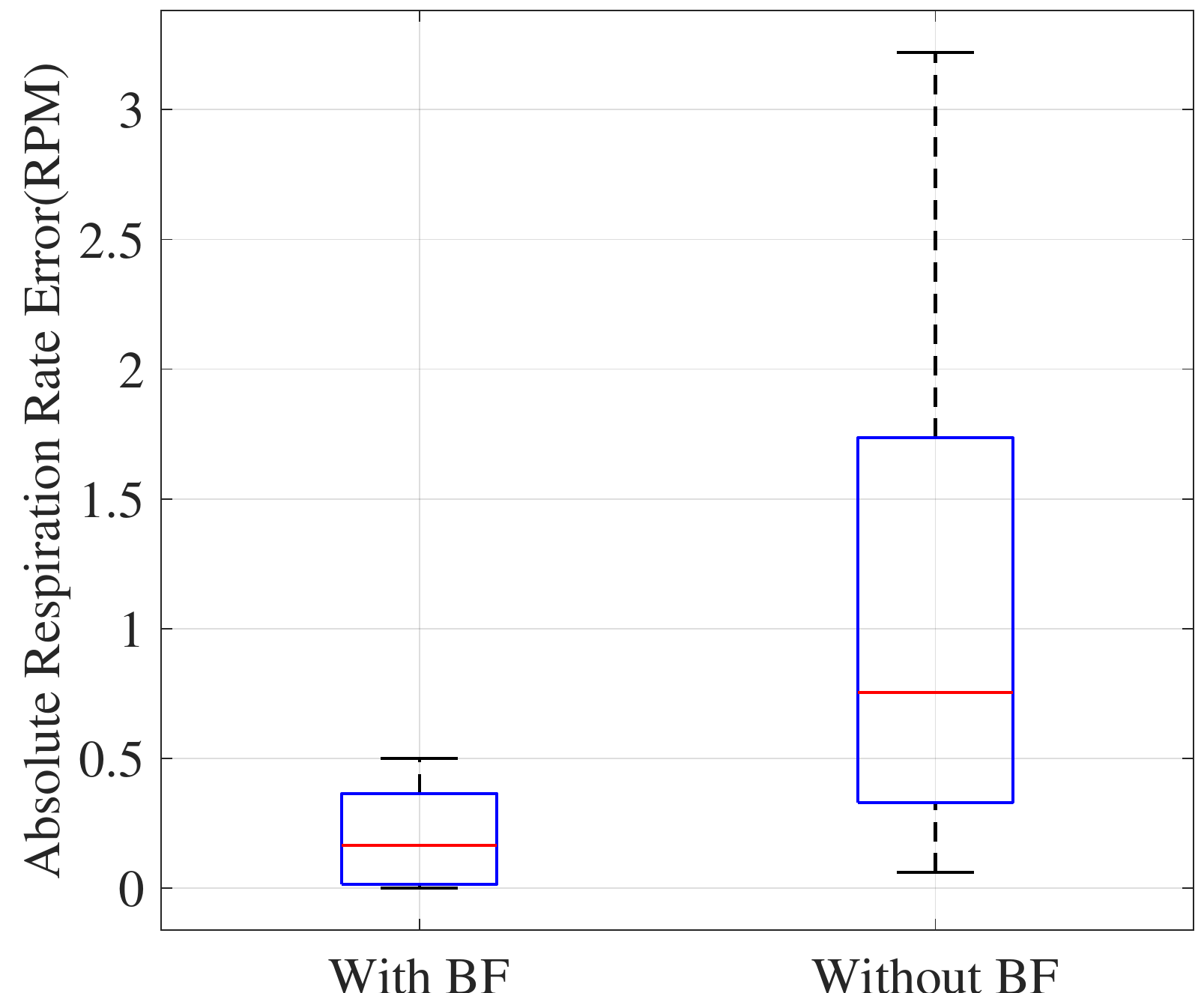}
    		\subcaption{Respiration}
    		\label{fig:18a}
    	\end{minipage}
    	\begin{minipage}[t]{0.49\linewidth}
    		\centering
    		\includegraphics[width=1\textwidth]{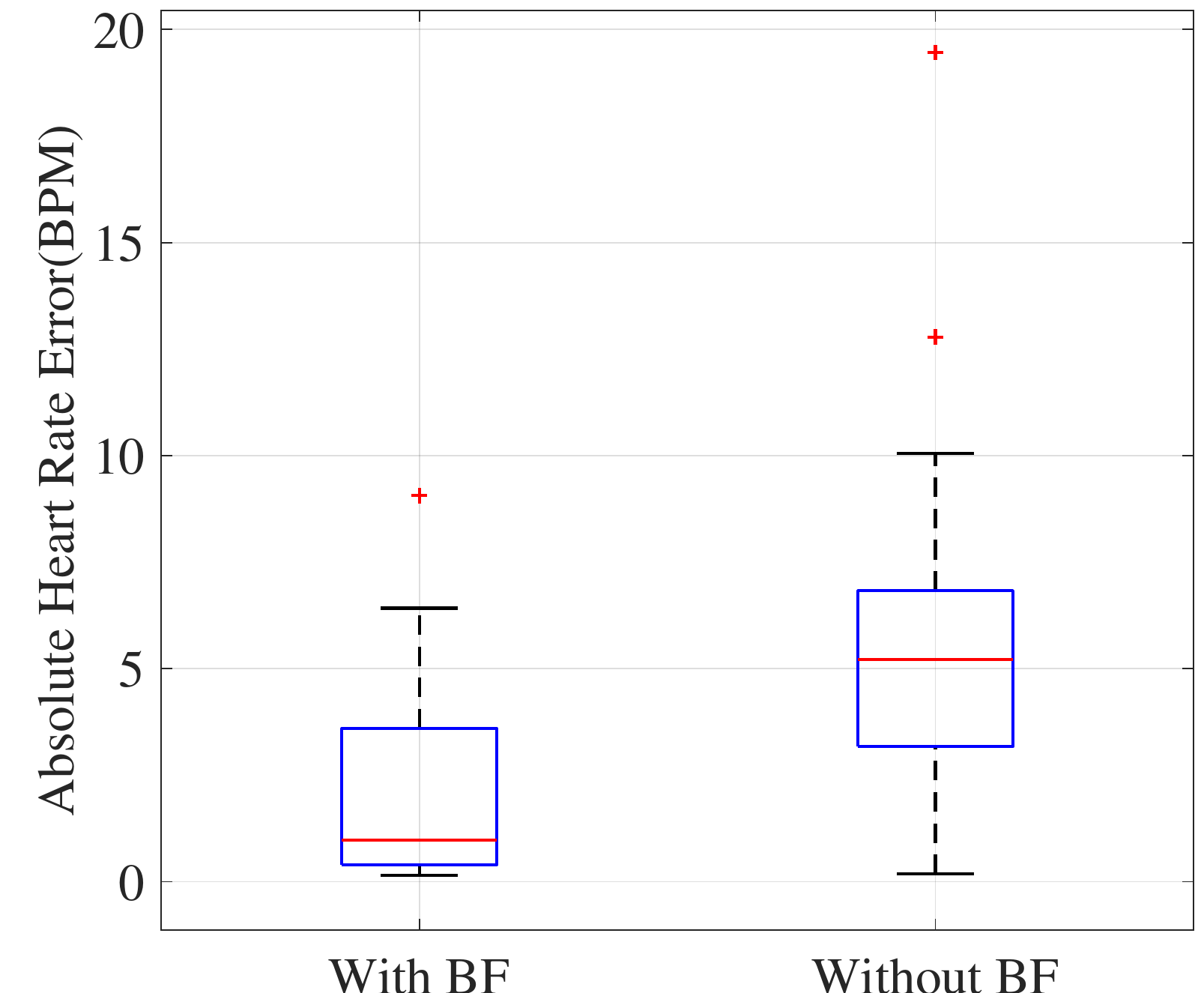}
    		\subcaption{Heart}
    		\label{fig:18b}
    	\end{minipage}
    	\vspace{-0.8em}
    	\caption{Impact of Beamforming}
    	\label{fig:18}
    \end{figure}

	In order to validate the effect of beamforming, we design an experiment to evaluate whether the signal from the target can be extracted validly under dynamic interference. In this experiment, two people are in front of the radar. In Fig. \ref{fig9a}, one is the target person who keeps still, while another person keeps moving near the target and plays the role of an interferer. Both of them are located at the same distance (2 m) but at different angles. The AoAs of the target and interferer are in the directions of $30 \degree$ and $60 \degree$ from the mmWave radar, respectively. Besides, the target person breathes normally while the disruptor shakes the arms at a high speed to disturb the target.
	
	Fig. \ref{fig9b} illustrates the filtered and normalized signal from these two persons using Tx and Rx beamforming. The blue line is the vital sign signal from the static person who is breathing stably. It can be observed that the waveform is periodic and superimposed by heartbeat. Another line shows the received signal collected from the interferer. Its frequency is higher while the signal amplitude is much weaker than the static target. This is contributed by Tx and Rx beamforming, which can amplify the signal from the direction of the target and attenuate the noise from the interferer. In practice, the signal amplitude caused by arm shaking should be much larger than the vital sign signal since the chest motion is very small. The result is in accordance with our expectations shown in Fig. \ref{fig:6d}.
	
	To further demonstrate the efficiency of beamforming, we compare the performance between enabling and disabling Tx- and Rx beamforming. One person is asked to walk around the target to induce interference. Moving closer to the target person is allowed. As shown in Fig. \ref{fig:18}, when BF is enabled, both the heart and respiration rates can achieve much lower errors. The RR error is 0.17 RPM with beamforming but it reaches 0.76 RPM without beamforming. For HR estimation, the error without BF is 5.21 BPM, over 5 times higher than the beamforming-enabled mode.

	
	\subsection{Impact of Body Motion}
	\begin{figure*}[h]
		\centering
		\begin{minipage}{0.35\textwidth}
			\centering
			\includegraphics[width=1\textwidth]{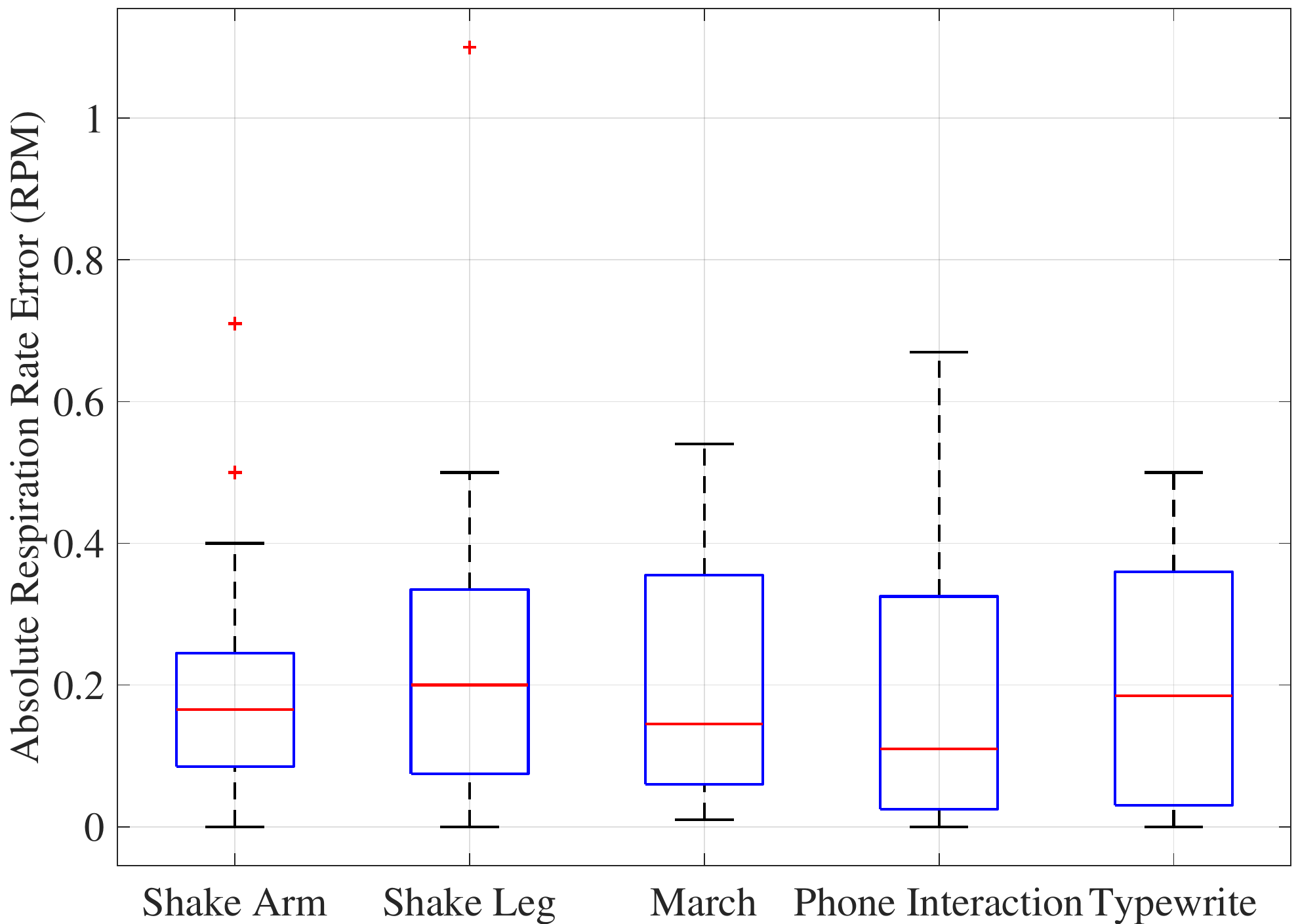}
			\subcaption{Respiration}
			\label{Fig14a}
		\end{minipage}
		\begin{minipage}{0.35\textwidth}
			\centering
			\includegraphics[width=1\textwidth]{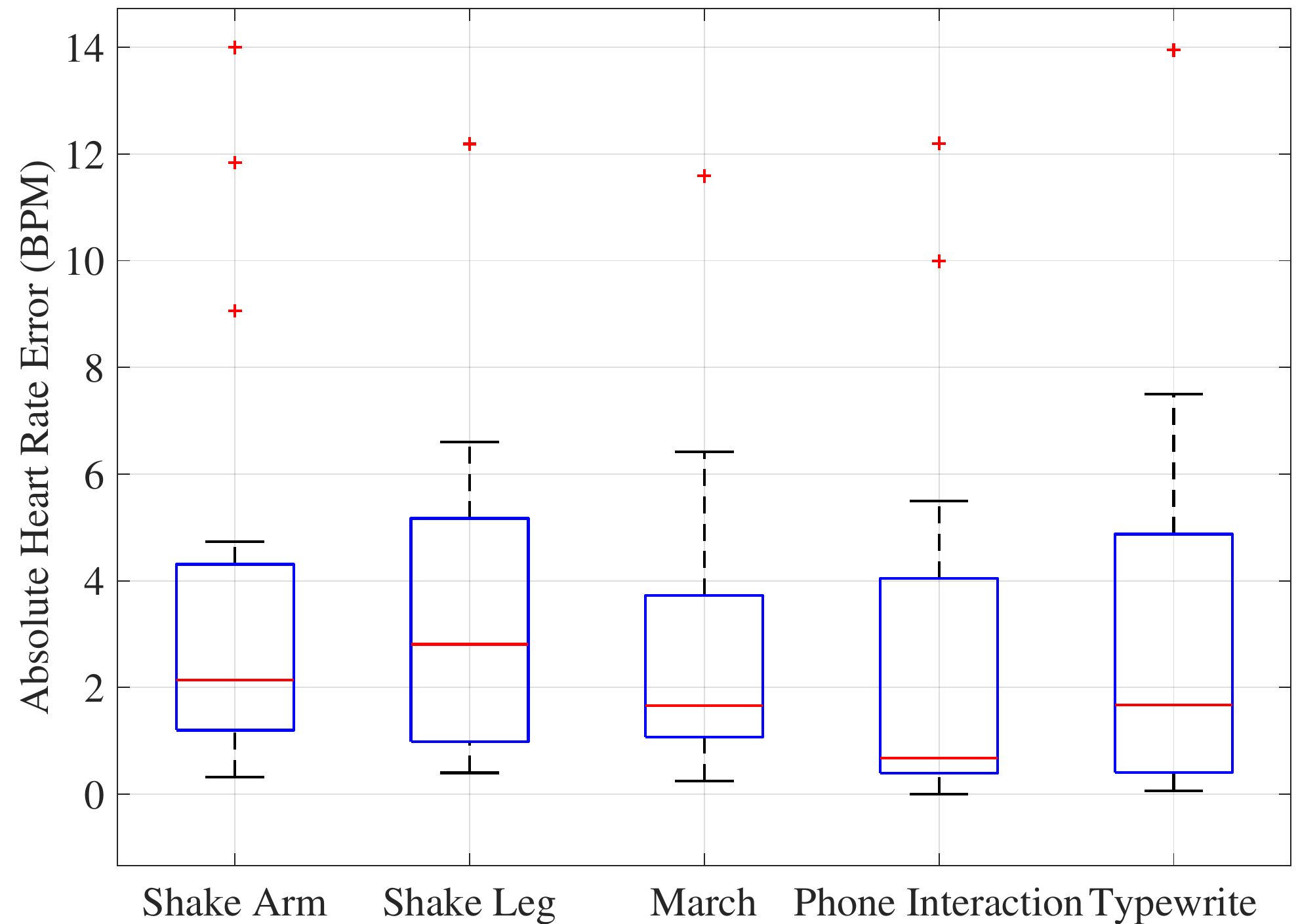}
			\subcaption{Heart}
			\label{Fig14b}
		\end{minipage}
		
		\begin{minipage}{0.19\textwidth}
			\centering
			\includegraphics[width=1\textwidth]{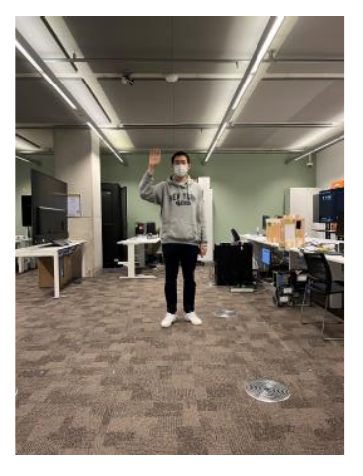}
			\subcaption{Shaking arm}
			\label{ges1}
		\end{minipage}
		\begin{minipage}{0.19\textwidth}
			\centering
			\includegraphics[width=1\textwidth]{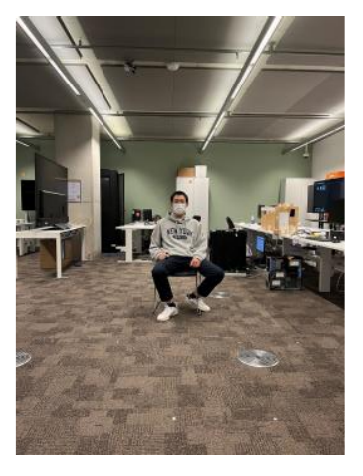}
			\subcaption{Shaking leg}
			\label{ges2}
		\end{minipage}
		\begin{minipage}{0.19\textwidth}
			\centering
			\includegraphics[width=1\textwidth]{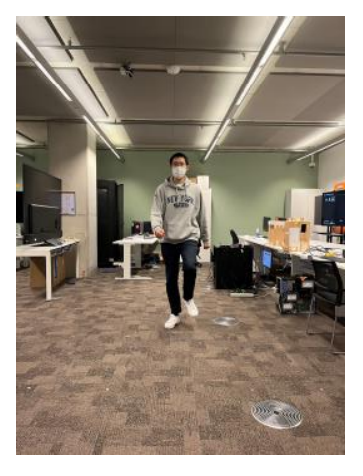}
			\subcaption{Marching}
			\label{ges3}
		\end{minipage}
		\begin{minipage}{0.19\textwidth}
			\centering
			\includegraphics[width=1\textwidth]{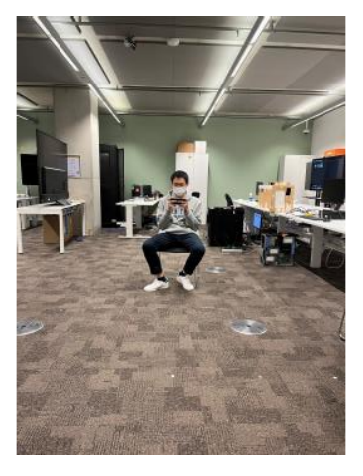}
			\subcaption{Phone interaction}
			\label{ges4}
		\end{minipage}
		\begin{minipage}{0.19\textwidth}
			\centering
			\includegraphics[width=1\textwidth]{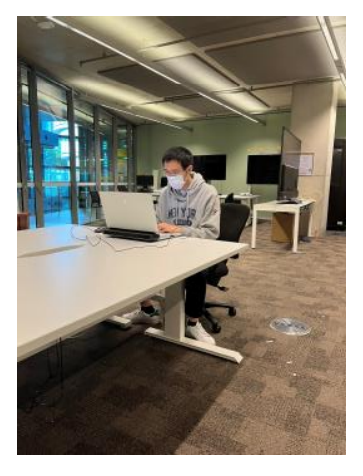}
			\subcaption{Typewriting}
			\label{ges5}
		\end{minipage}
		
		\vspace{-0.8em}
		\caption{Impact of body motion}
		\label{Fig14}
		\vspace{-1.5em}
	\end{figure*}
	
	To investigate the impact of different motion gestures, we ask the tracked person to perform five types of daily motions at the range of 3 m to the mmWave radar: shaking arm, shaking leg, march, phone interaction and typewriting. Specifically, the person waves their arms and shakes leg and this lasts in the first two motion types. Both arms and legs move periodically over the whole sampling period in march. In the gesture of phone interaction, the person uses the cellphone to do their intended interaction, such as typing messages, browsing information and playing games. This motion mainly contains the finger motion. Finally, the person adjusts their body gesture during typewriting. This gesture includes the random motion of fingers, arms and torso.
	
	The result is shown in Fig. \ref{Fig14}. The trends of respiration and heart rates error are similar. When the target person performs phone interaction, both HR and RR reach the lowest error (0.11 RPM and 0.68 BPM). Shaking leg contributes to the highest error rates (0.2 RPM and 2.81 BPM). The error rates are similar in the rest of the gestures but they can still adversely affect the accuracy.	
	
	From the results, we have the following observations: \textbf{1) Phone interaction.} Since interacting with cellphone only contains the motion of fingers and the extent is much smaller than other types of motions, it has the lowest impact on the estimation accuracy. 2) \textbf{Shaking leg.} The frequency of the shaking leg is very close to the heartbeat, which interferes with estimation of heart rate. \textbf{3) Typewriting.} Typewriting affects more in respiration rate because there could be arm and torso motion with a close frequency to respiration rate. \textbf{4) Shaking arm.} Shaking arm is of the greatest extent. This contributes to both large errors in RR and HR. \textbf{5) Marching.} On the contrary, marching is performed over the sample period and the frequency is close to the heartbeat. This leads to a higher error rate in heart rate estimation but has less impact on respiration rate estimation.

	\subsection{Impact of Moving People}
	\begin{figure*}
		\centering
		\begin{minipage}[t]{0.329\linewidth}
			\centering
			\includegraphics[width=1\textwidth]{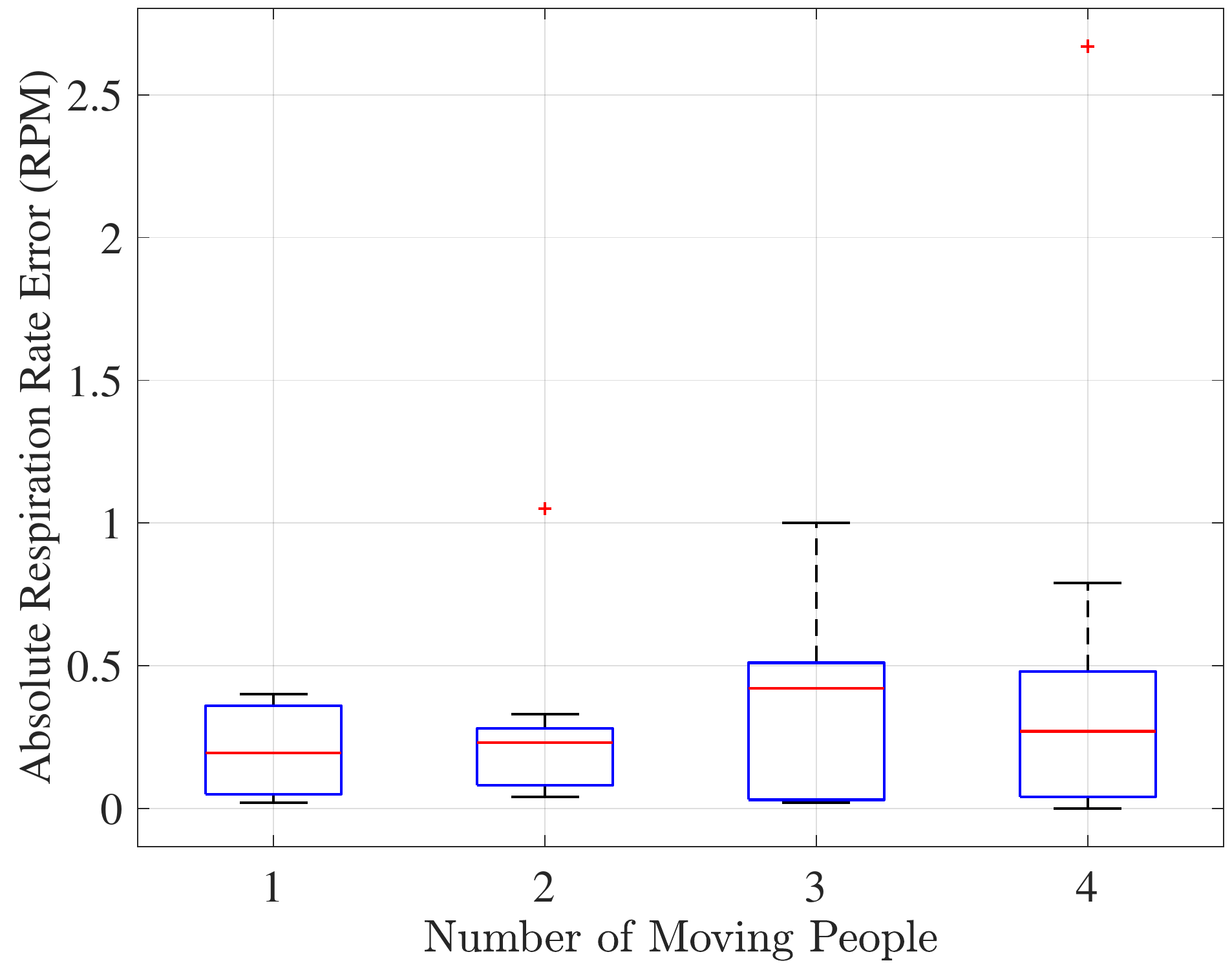}
			\subcaption{Respiration}
			\label{fig:15a}
		\end{minipage}
		\begin{minipage}[t]{0.329\linewidth}
			\centering
			\includegraphics[width=1\textwidth]{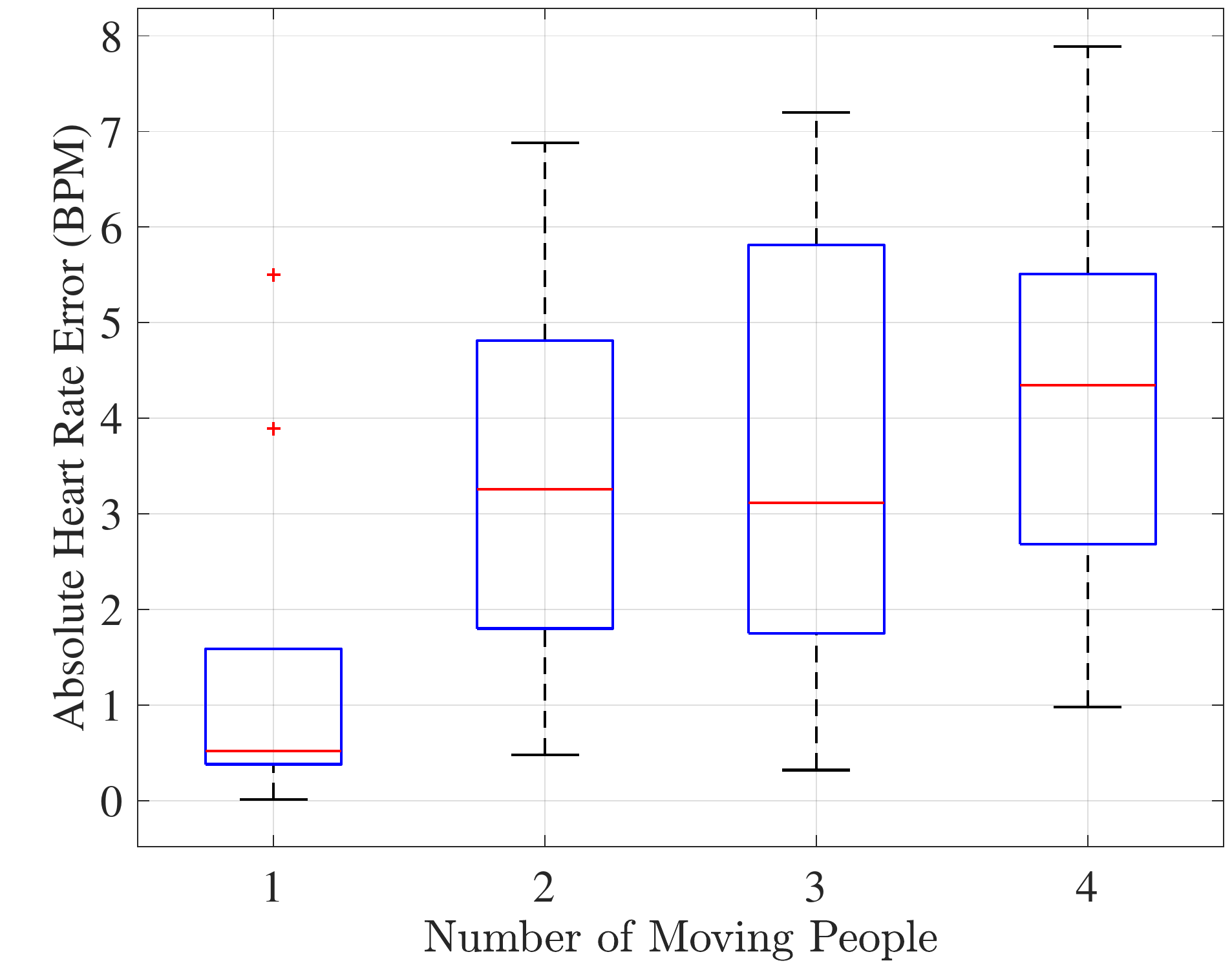}
			\subcaption{Heart}
			\label{fig:15b}
		\end{minipage}
		\begin{minipage}[t]{0.329\linewidth}
			\centering
			\includegraphics[width=0.9\textwidth]{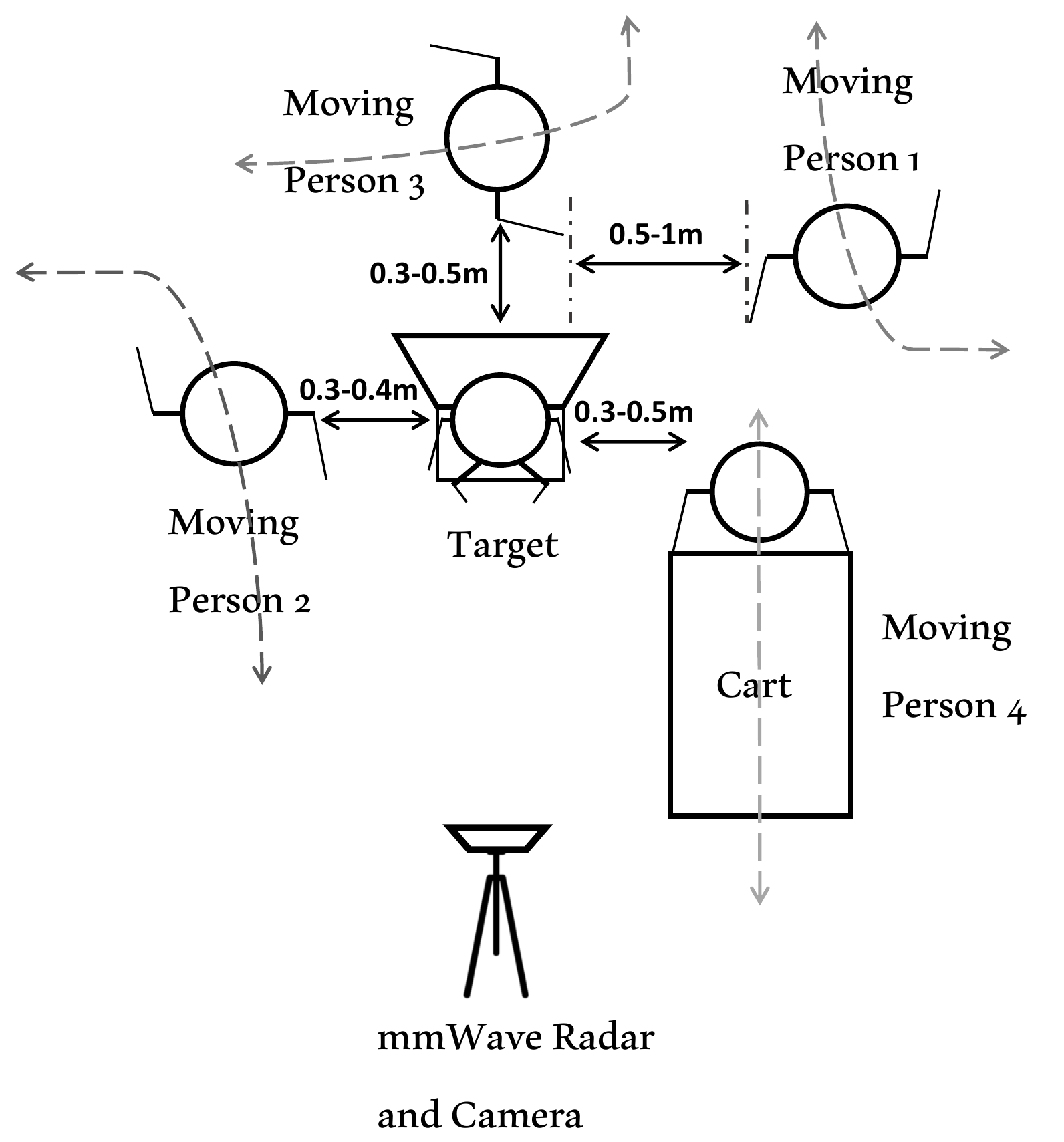}
			\subcaption{Bird's view of multiple interferers. The interferers move repetitively along the routes. Same routes and spots are also set in 1, 2 and 3 interferers cases.}
			\label{fig:15d}
		\end{minipage}
		
		\begin{minipage}[t]{1\linewidth}
			\centering
			\includegraphics[width=1\textwidth]{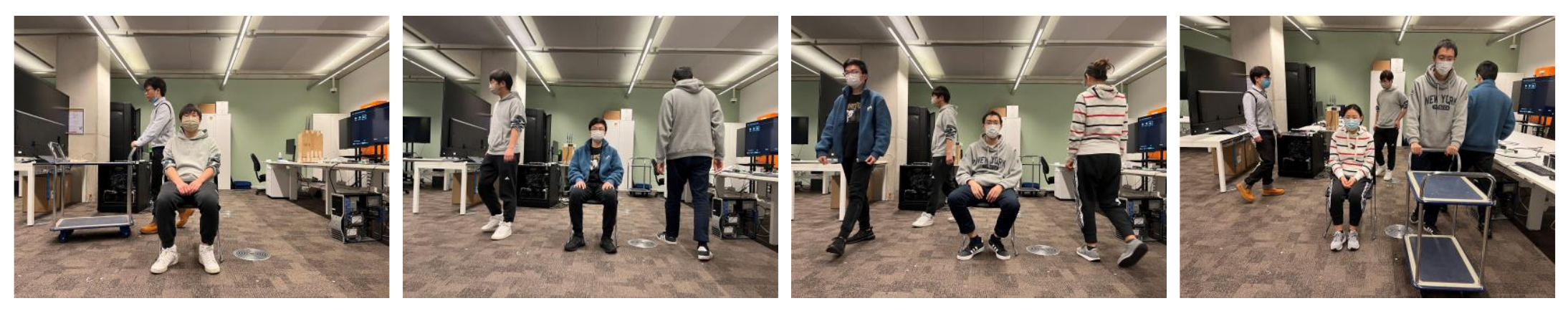}
			\subcaption{Different Moving People Scenarios. In condition of four moving people, a cart is pushed by a person to induce more interference.}
			\label{fig:15c}
		\end{minipage}

		\vspace{-0.8em}
		\caption{Impact of Moving people}
		\label{fig:15}
	\end{figure*}
	
	
	To study the impact of dynamic interference, four volunteers are asked to move in space. Fig. \ref{fig:15c} shows the testing scenarios with one to four moving persons. In each experiment, each person is asked to walk past the target by turns. 
	
	Fig. \ref{fig:15a} and \ref{fig:15b} show the impact of moving people. The figures demonstrate that the error coarsely increases with increased moving individuals. We can see that the heart rate is easier to be interfered by moving people while respiration rate estimation is less affected. Our system can achieve the lowest error when there is one moving person (0.195 RPM for respiration rate and 0.52 BPM for heart rate). The respiration error reaches the highest (i.e., 0.42 RPM) in the three-people scenario. The
	highest error of heartbeat estimation is 4.35 BPM in the four-people scenario.
	
	
	
	\subsection{Impact of Environment}
	\begin{figure*}
		\centering
		\begin{minipage}[t]{0.29\linewidth}
			\centering
			\includegraphics[width=1\textwidth]{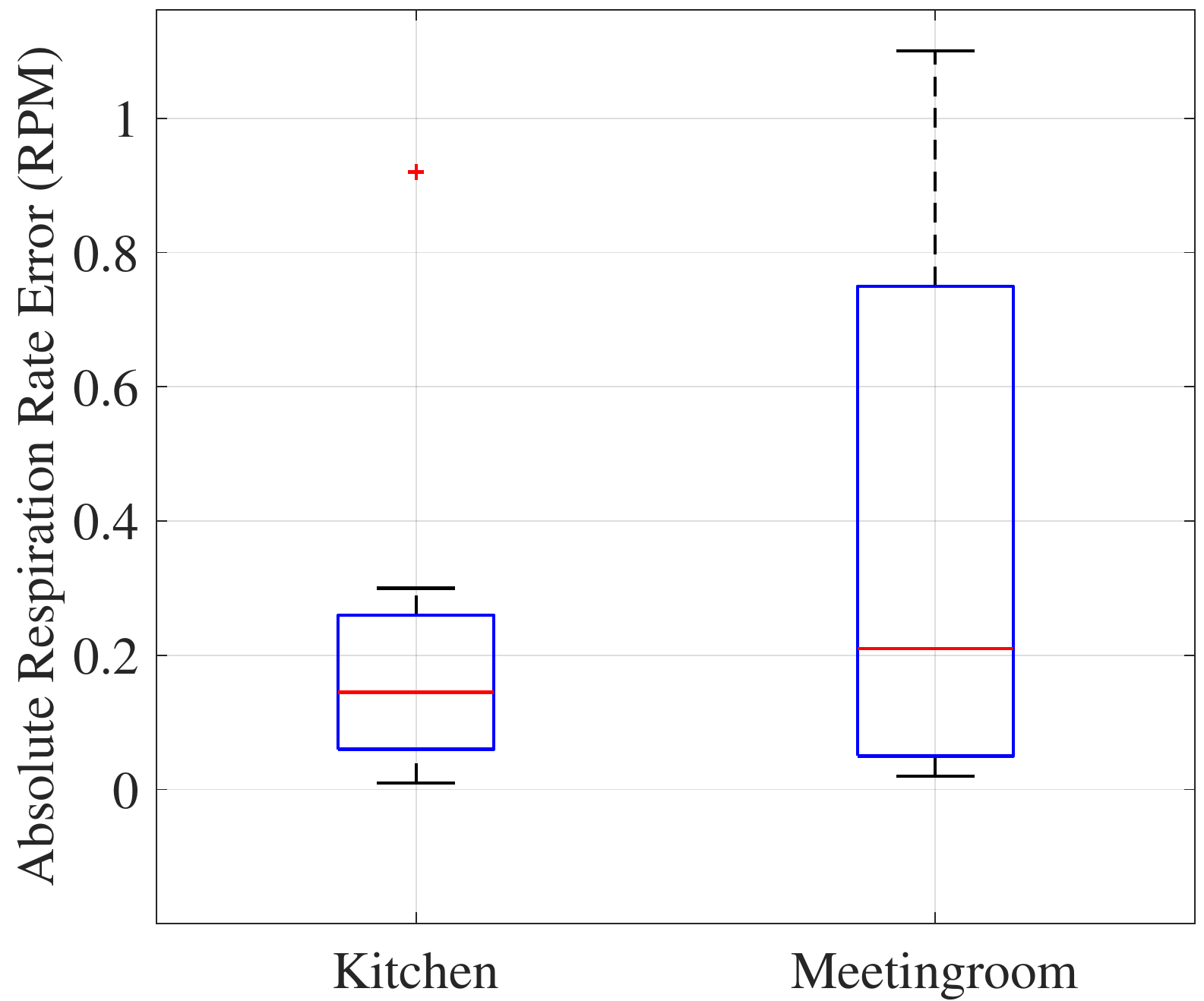}
			\subcaption{Respiration}
			\label{fig:16a}
		\end{minipage}
		\begin{minipage}[t]{0.29\linewidth}
			\centering
			\includegraphics[width=1\textwidth]{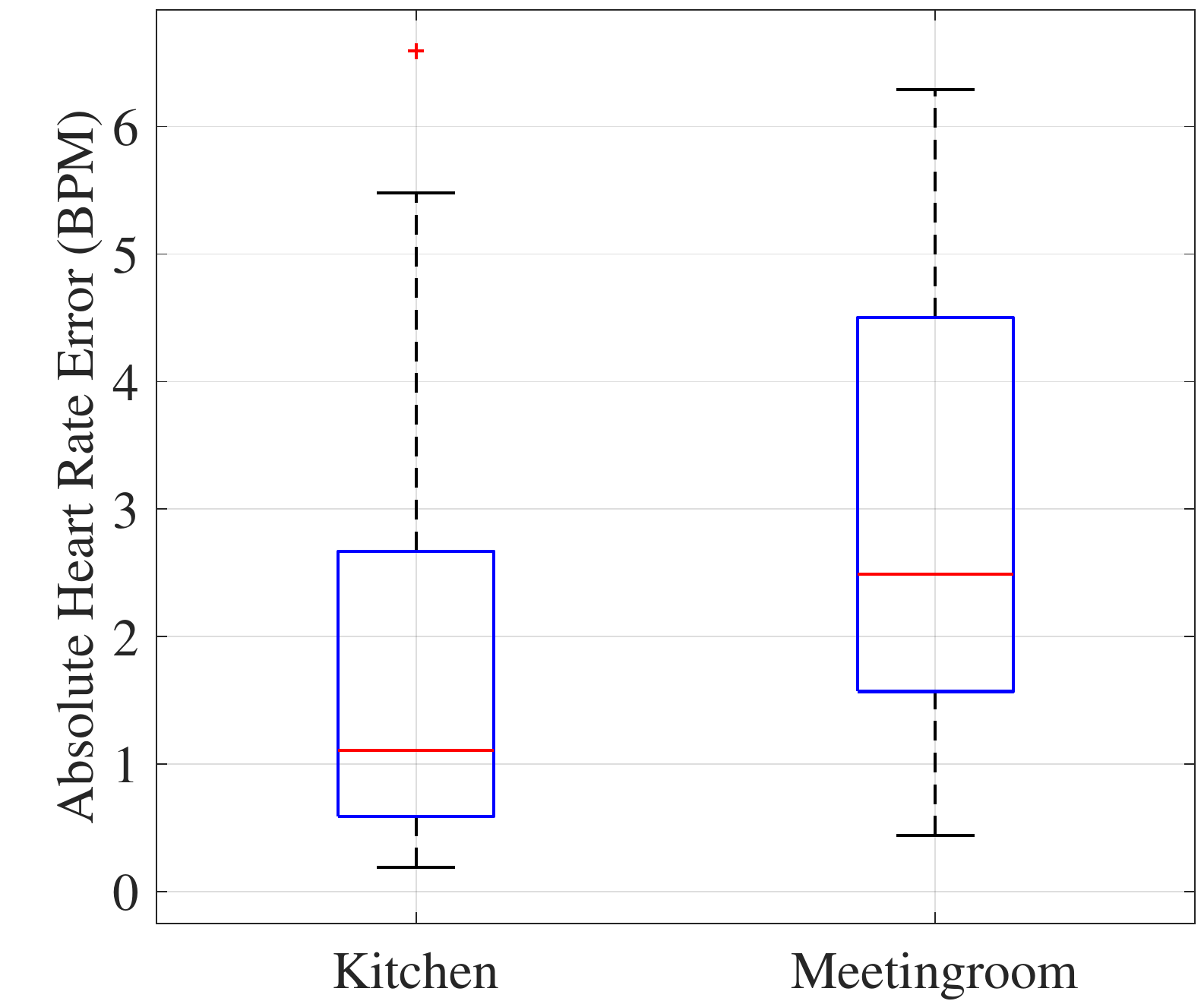}
			\subcaption{Heart}
			\label{fig:16b}
		\end{minipage}
		\begin{minipage}[t]{0.4\linewidth}
			\centering
			\includegraphics[width=1\textwidth, height=4.5cm]{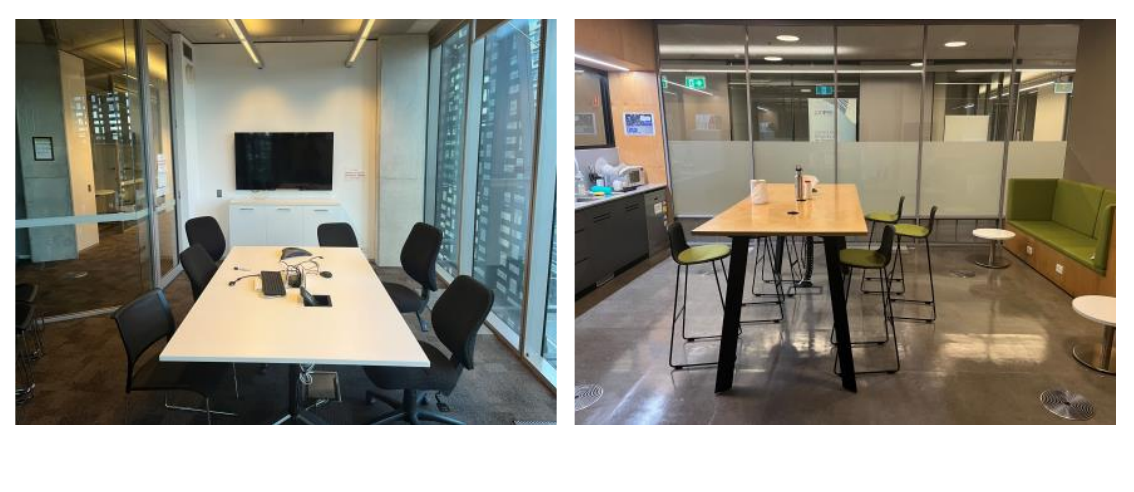}
			\subcaption{Meeting room and kitchen}
			\label{rest}
		\end{minipage}
		\vspace{-0.8em}
		\caption{Impact of Environments}
		\label{fig:16}
	\end{figure*}
	
	We also conducted the experiment in a meeting room and restaurant (shown in Fig. \ref{rest}) to evaluate the impact of the environment. In both two scenarios, the different individuals are allowed to enter and leave the room at any time. Also, the testing target is asked to randomly perform the intended gestures, such as typewriting in the meeting room and eating or drinking in the kitchen. 
	
	As shown in Fig. \ref{fig:16a}, \ref{fig:16b}, the accuracy of heart and respiration rates in the kitchen is always higher in comparison to the meeting room. This could be related to the size of these two places. The meeting room is more crowded. It implies that more interfering signals from passengers' movement are superimposed on the vital sign signal due to the short distance during walking. 
	
	\subsection{Impact of Range}
	\begin{figure*}
		\centering
		\begin{minipage}[t]{0.329\linewidth}
			\centering
			\includegraphics[width=1\textwidth]{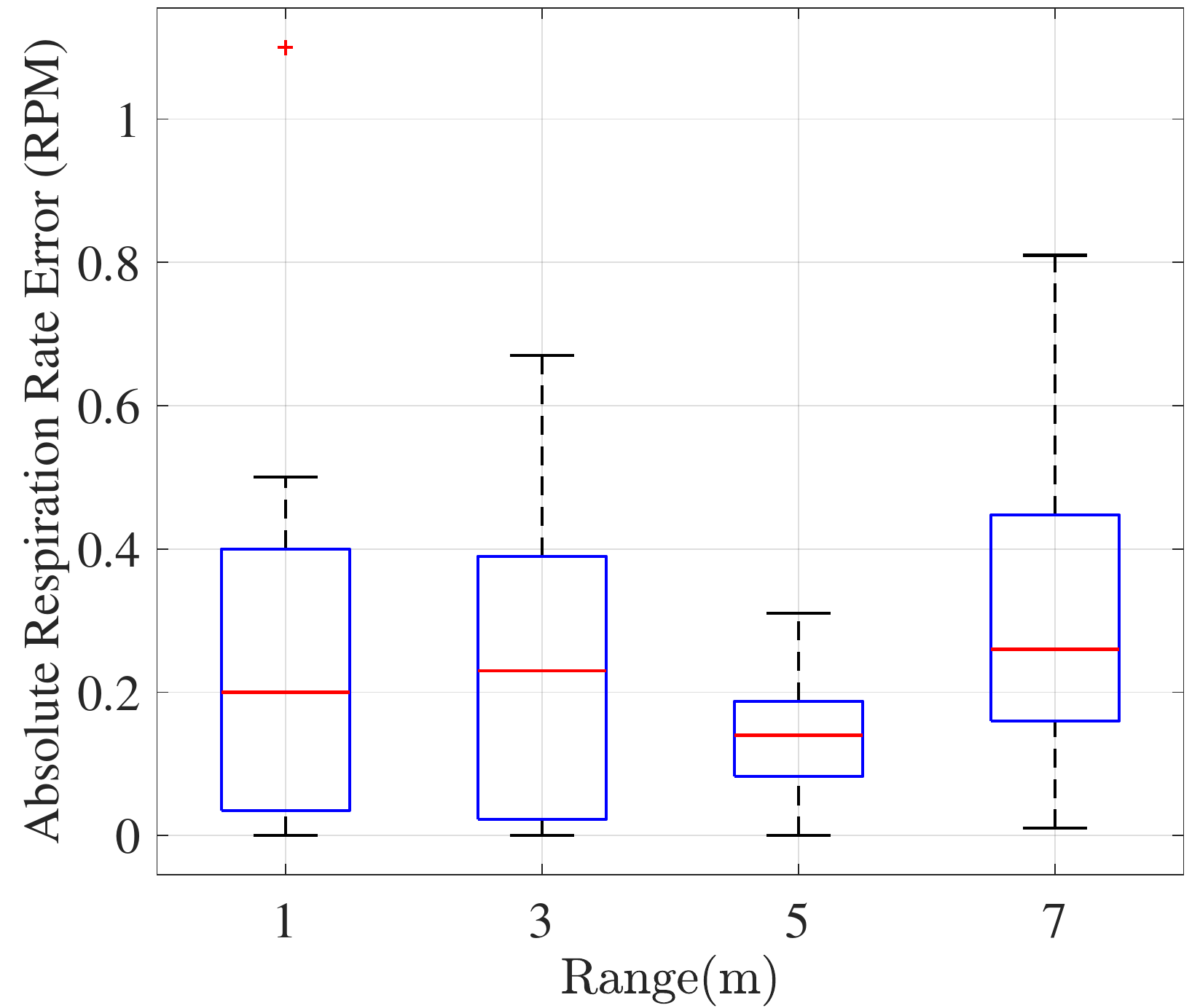}
			\subcaption{Respiration}
			\label{fig:13a}
		\end{minipage}
		\begin{minipage}[t]{0.329\linewidth}
			\centering
			\includegraphics[width=1\textwidth]{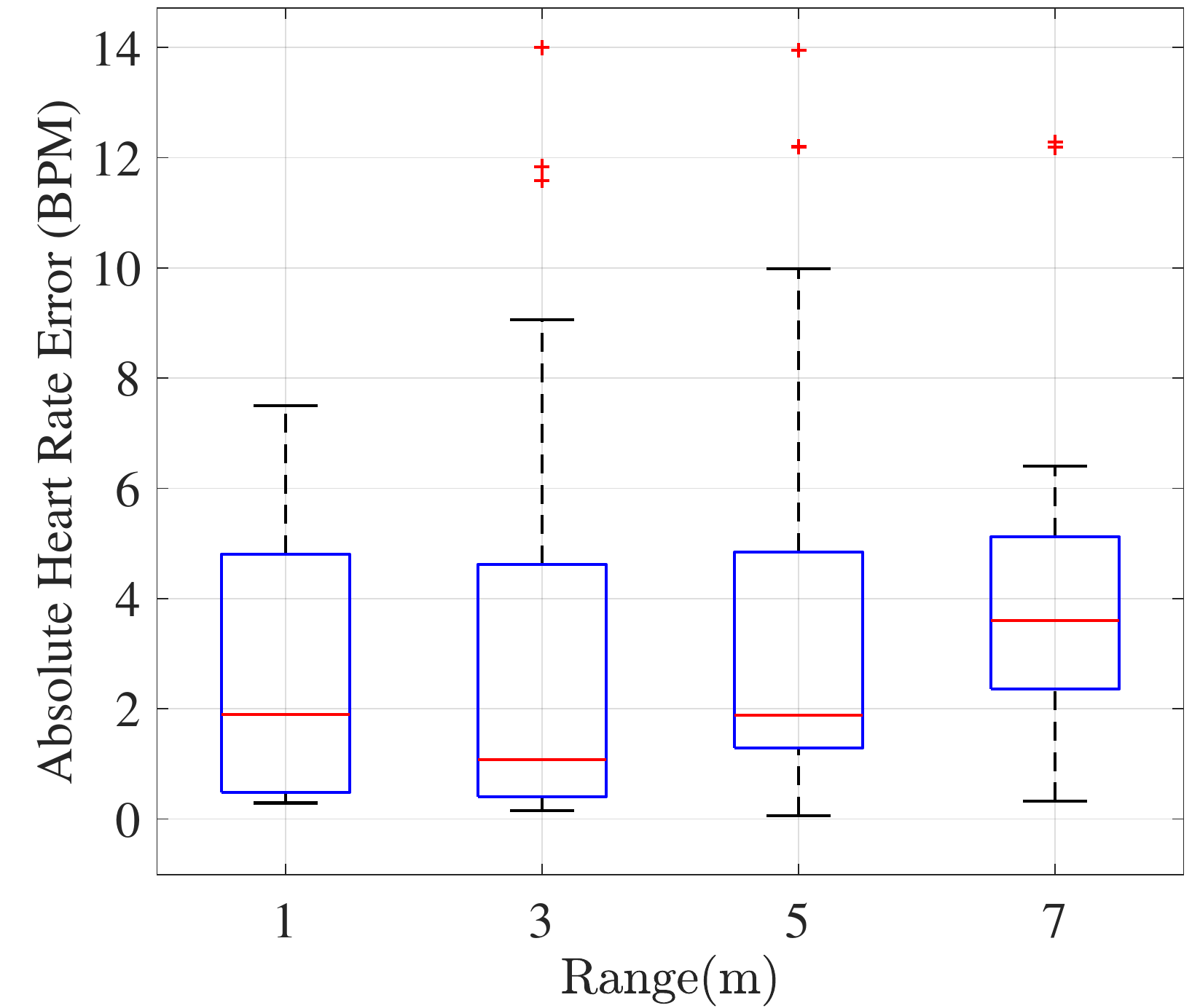}
			\subcaption{Heart}
			\label{fig:13b}
		\end{minipage}
		\begin{minipage}[t]{0.329\linewidth}
			\centering
			\includegraphics[width=0.75\textwidth, height=5cm]{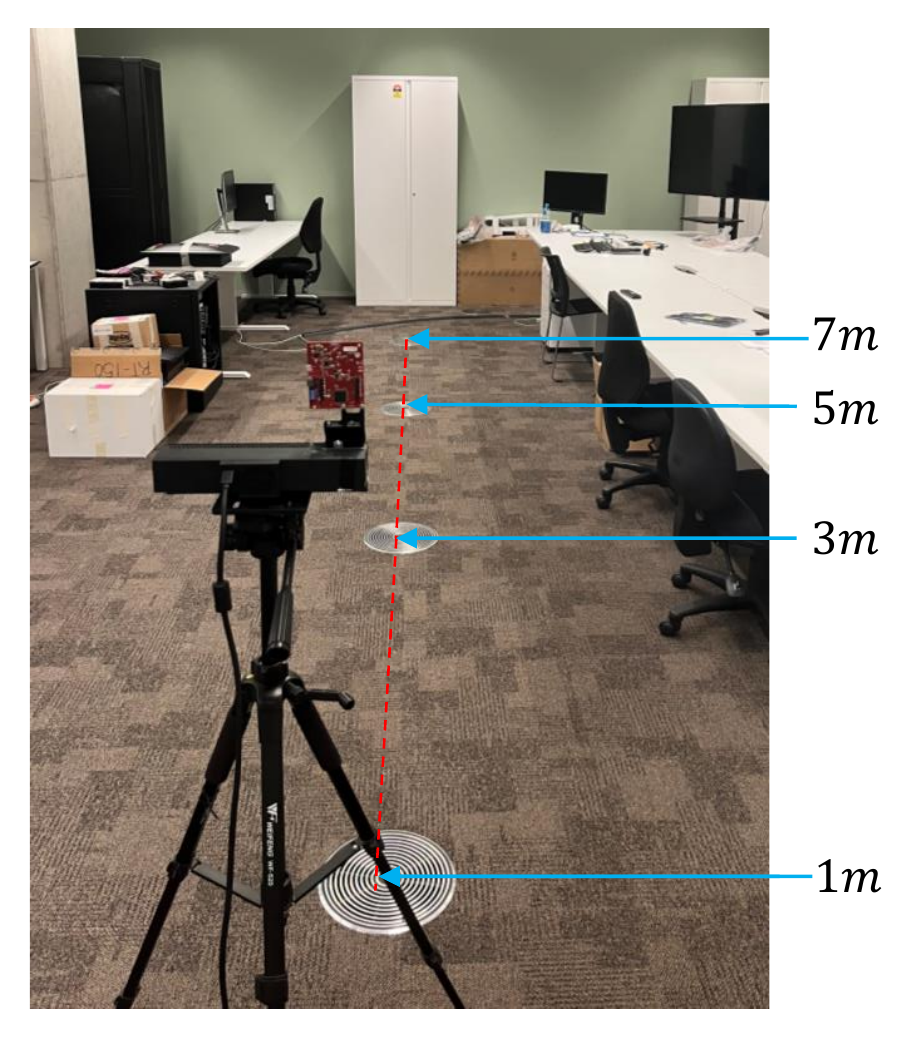}
			\subcaption{Range}
			\label{fig:13c}
		\end{minipage}
		\vspace{-0.8em}
		\caption{Impact of different range. The plates on the ground are placed in 1, 3, 5 and 7m as the signs of different range.}
		\label{fig:13}
	\end{figure*}
	
	This experiment verifies the accuracy of the heart and respiration rate estimated at different distances to the mmWave radar, i.e., 1m, 3m, 5m and 7m (shown in Fig. \ref{fig:13c}). In each experiment, the target person repeats the same gestures and the interfering persons along the same routine. The experiment results are shown in Fig: \ref{fig:13a} and \ref{fig:13b}.
	
	For the respiration rate, the accuracy goes down when the range increases. The minimum and maximum errors appear at the range of 1m and 7m, respectively. The error at 7 m is 0.26 RPM which is nearly double as the error (0.14 RPM) at 1 m. The heart rate is more sensitive to the variation of range. The maximum error is 3.6 BPM, which is more than three times the minimum error rate (1.08 BPM) at 3 m. The results reveal that when the range increases, the signal will be intensively attenuated. In this case, the weaker vital sign signal is more sensitive to interferences.
	
	\subsection{Impact of Acceleration scheme}
	\begin{table}
		\caption{Acceleration and Without Acceleration performance}
		\vspace{-1.5em}
		\label{table0}
		\center
		\begin{tabular}{c|c|c|c}
			\hline
			& RR (RPM) & HR (BPM) & Time (ms) \\ \hline
			Without Acceleration & 0.33        & 5.02        & 25.35         \\ \hline
			With Acceleration    & 0.29        & 3.06        & 8.56         \\ \hline
		\end{tabular}
		\vspace{-2em}
	\end{table}
	We also evaluate the performance of our proposed acceleration strategy. Specifically, we choose the first 100 components on the whole spectrum to estimate the RR and HR. And we compare the error and time consumption between the application and nonapplication of the acceleration scheme. As shown in Table 4, when the acceleration scheme is applied, the average computation time is 8.56 ms, which is approximately 1/3 of the one without acceleration (25.25 ms).

	
	It is surprising to see that the error of RR and HR (0.29 RPM and 3.06 BPM) with acceleration are both smaller than those in non-acceleration case (0.33 RPM and 5.02 BPM). The possible reason is as follow. The respiration and heartbeat signals are both low-frequency. When we choose all components in the spectrum, those high-frequency components often contribute as noise and adversely affect the decomposition. When we only choose the low-frequency components, those high-frequency noises are eliminated, which improves the decomposition accuracy.

	\subsection{Impact of Obstacles/Non Line of Sights}
\begin{figure*}[h]
		\centering
		\begin{minipage}[t]{0.327\linewidth}
			\centering
			\includegraphics[width=0.98\textwidth]{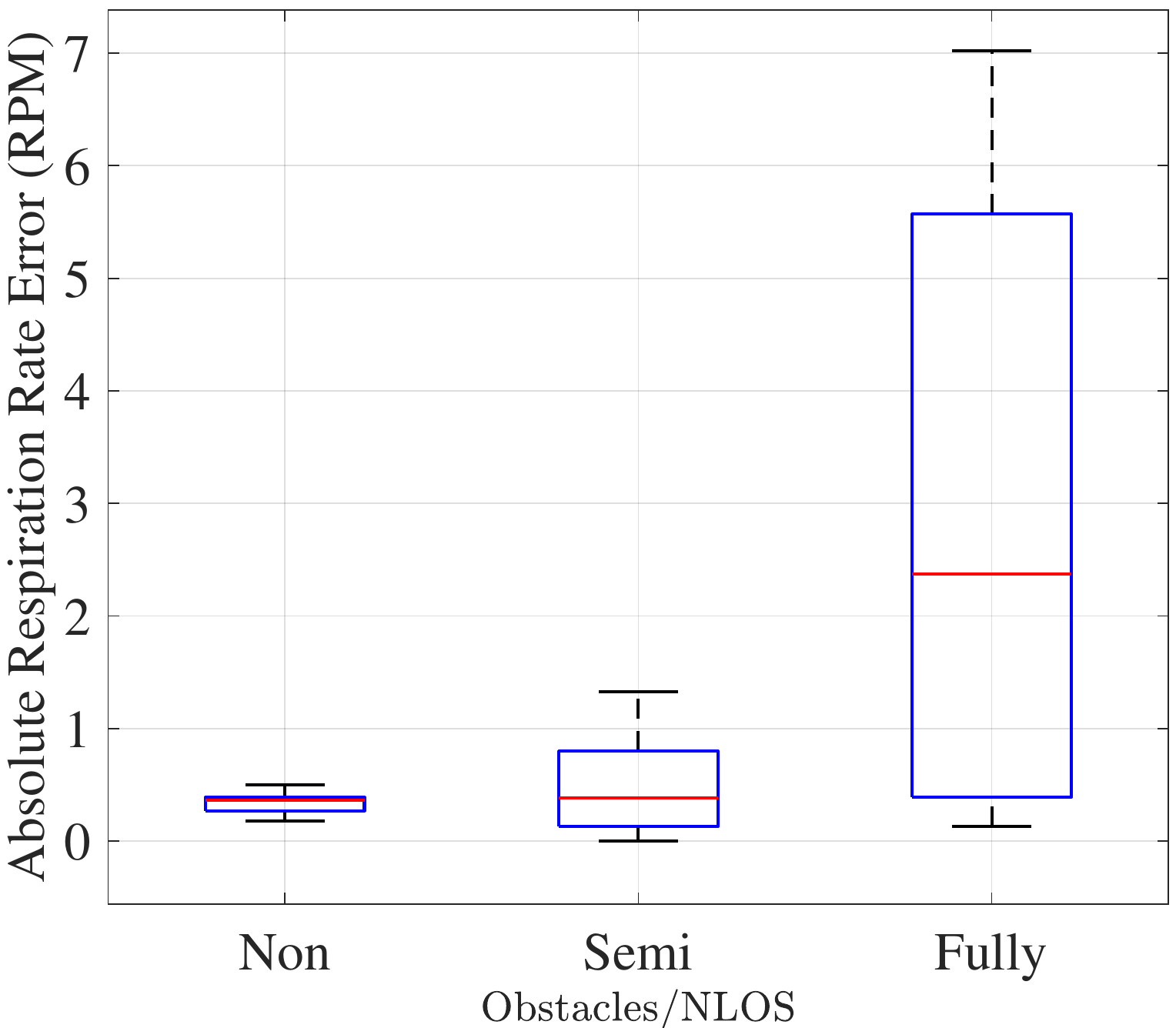}
			\subcaption{Respiration}
			\label{fig:17a}
		\end{minipage}
		\begin{minipage}[t]{0.327\linewidth}
			\centering
			\includegraphics[width=1\textwidth]{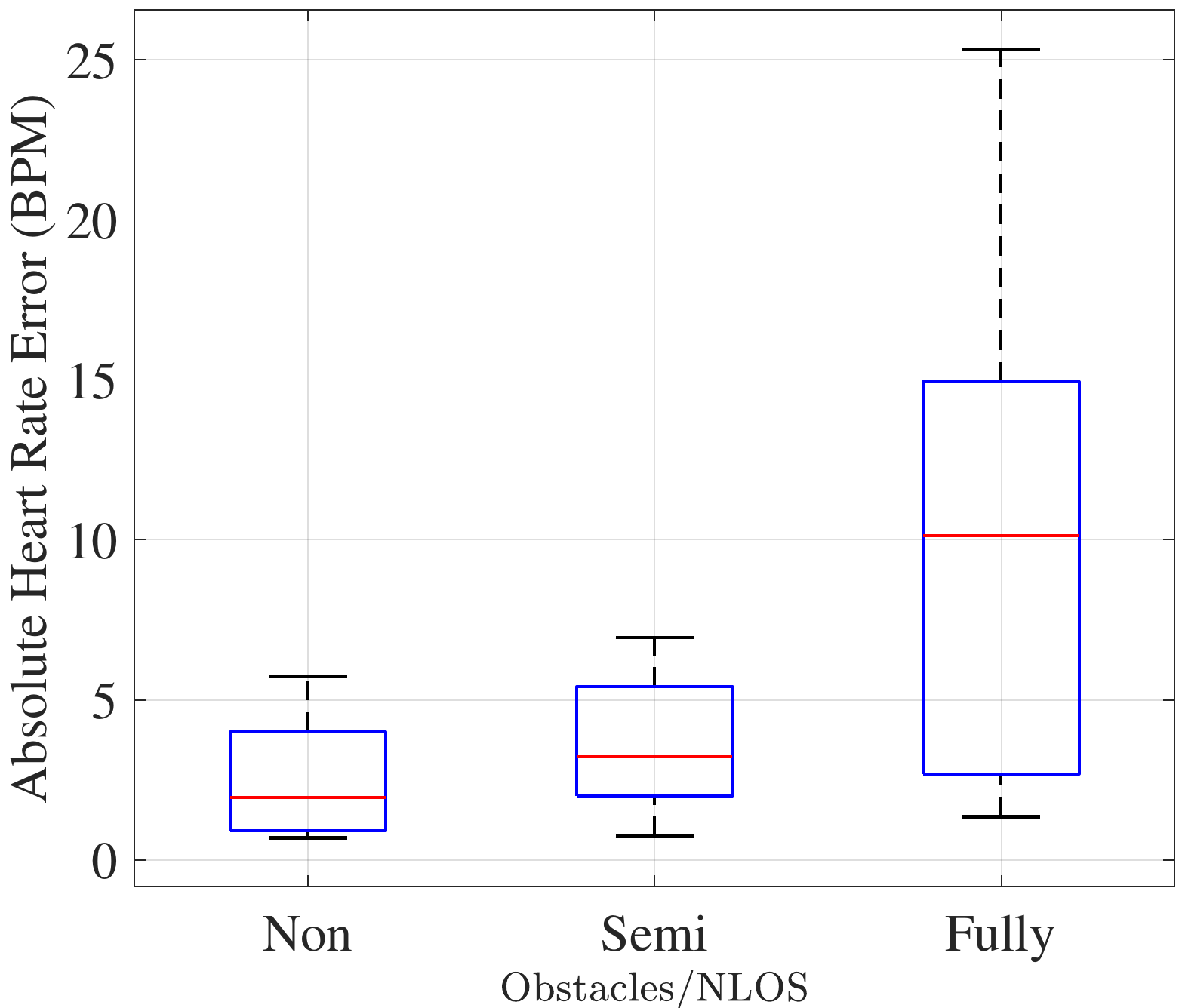}
			\subcaption{Heart}
			\label{fig:17b}
		\end{minipage}
		\begin{minipage}[t]{0.335\linewidth}
			\centering
			\includegraphics[width=1.1\textwidth, height=4.25cm]{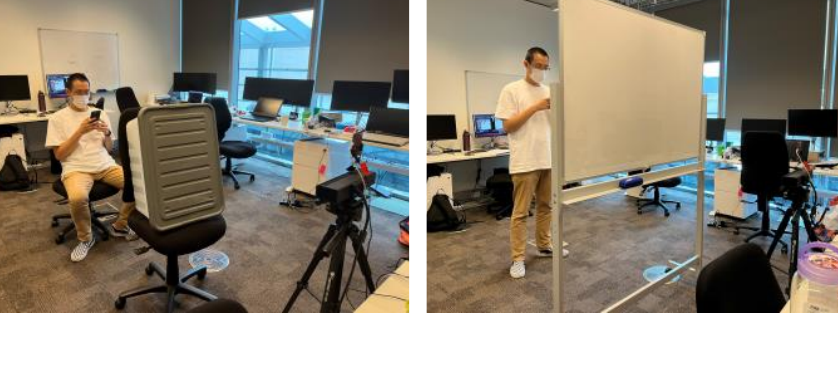}
			\subcaption{Left: semi-sheltered, Right: Fully-sheltered}
			\label{fig:17c}
		\end{minipage}
		\vspace{-0.8em}
		\caption{Impact of Obstacles/NLoS}
		\label{fig:17}
	\end{figure*}

	It is also interesting to know how our scheme performs when there are some obstacles between target and radar. To evaluate the performance, we carry out two experiments for semi-sheltered and fully-sheltered obstacles respectively. In semi-sheltered case (shown in Fig. \ref{fig:17c} Left), a large-sized plastic tank is placed in front of the radar and the target is seated behind the tank. In the second case, the target is in the position behind a large and solid white board (shown in Fig. \ref{fig:17c} Right). The board is capable of covering the whole upper body. In the second case the target location is known in prior since the camera cannot capture the object in fully NLoS . But the Tx-Rx beamforming and WMC-VMD are still working to enhance the target's signal and suppress noise and alleviate body motions. The results are shown in Fig. \ref{fig:17a} and \ref{fig:17b}.
	
	As expected, the lowest error rates (0.365 RPM for RR and 1.96 for HR) appear when there is no obstacle. It can be observed that the error rates are still acceptable in the semi-sheltered case, and the RR error and HR error are still less than 0.5 RPM and 2 BPM. This implies that the radio signal could bypass the obstacle if its size is not too large or radar signal could penetrate the obstacle. In fully-sheltered case, despite of the low RR error rate (2.38 RPM), the HR error is beyond 10 BPM, which shows the scheme cannot work in fully-sheltered scenarios for HR monitoring but is still able to deal with RR monitoring. This is because the signal variation due to heart beating, which is much smaller than the variation of breathing, becomes too weak to be observed.
	
	\subsection{Impact of Target Orientation}
	\begin{figure*}[h]
		\centering
		\begin{minipage}[t]{0.329\linewidth}
			\centering
			\includegraphics[width=0.98\textwidth]{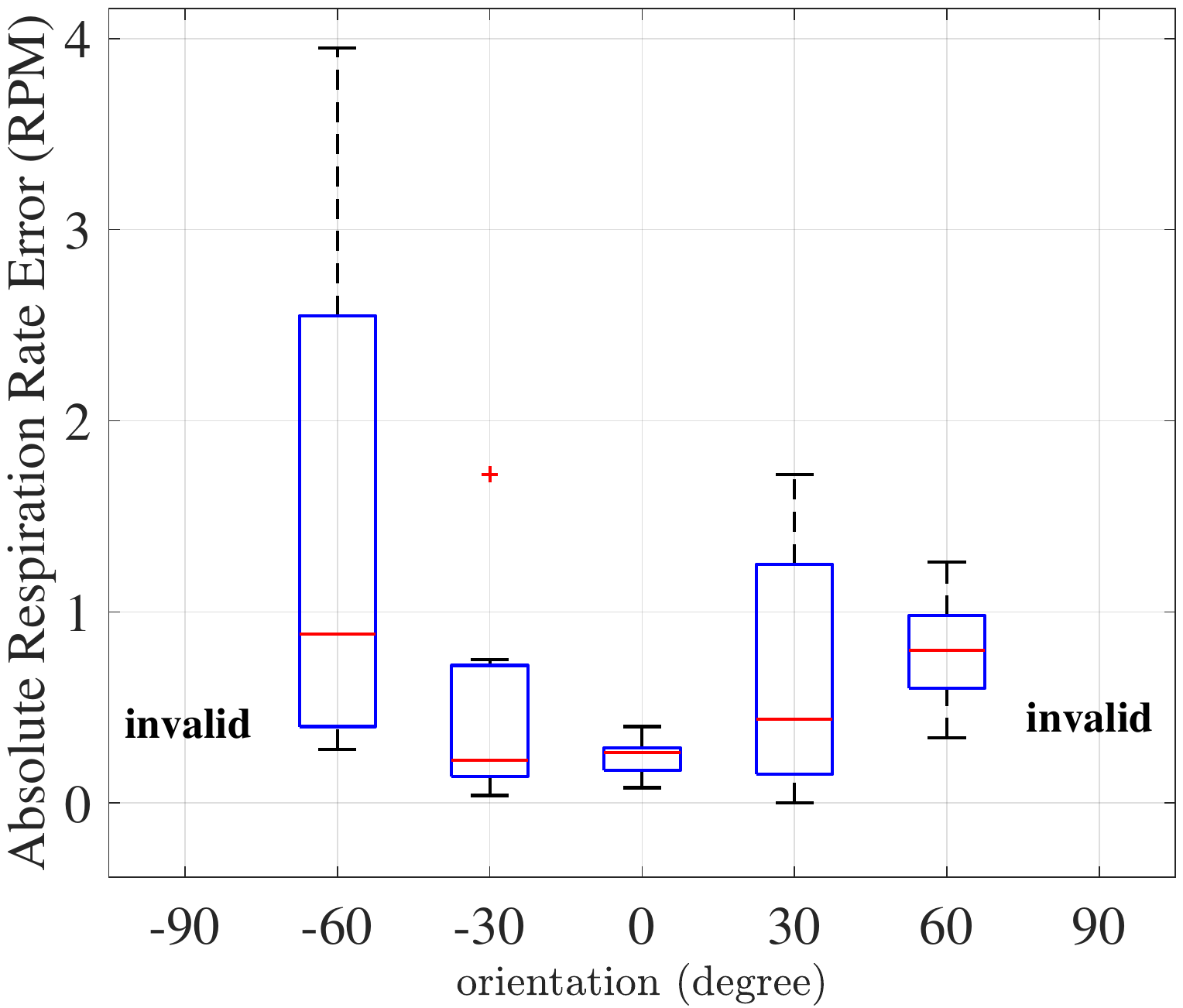}
			\subcaption{Respiration}
			\label{fig:19a}
		\end{minipage}
		\begin{minipage}[t]{0.329\linewidth}
			\centering
			\includegraphics[width=1\textwidth]{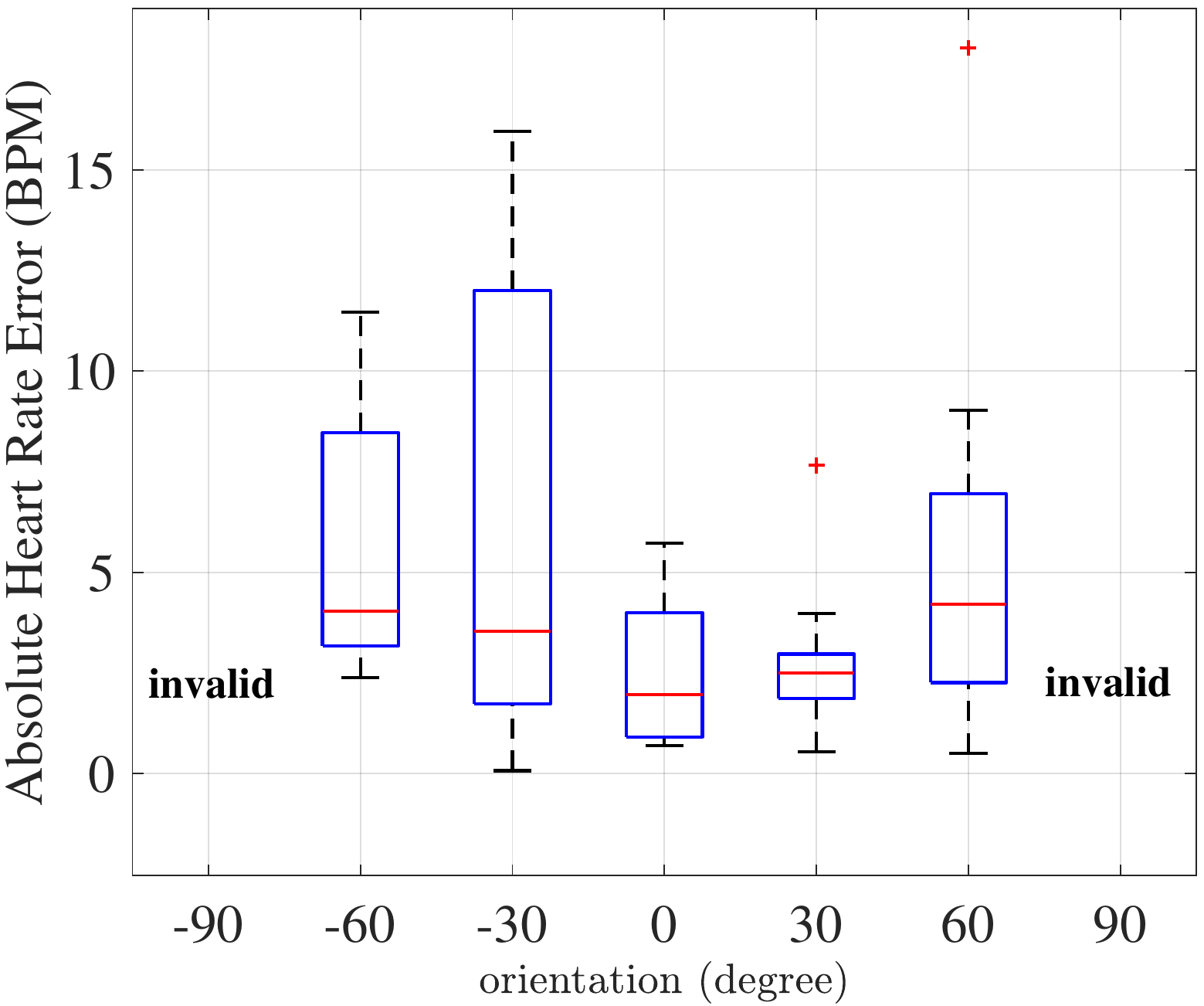}
			\subcaption{Heart}
			\label{fig:19b}
		\end{minipage}
		\begin{minipage}[t]{0.329\linewidth}
			\centering
			\includegraphics[width=0.95\textwidth, height=5.5cm]{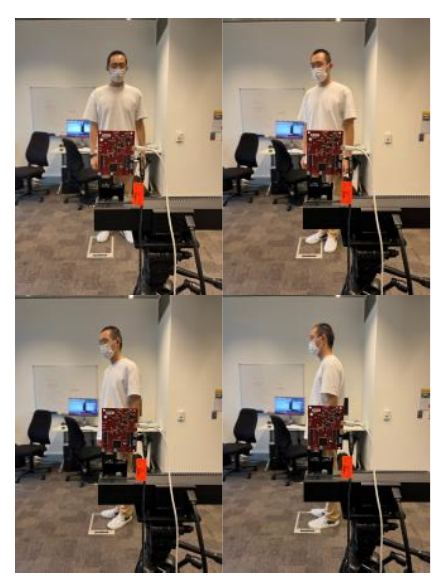}
			\subcaption{Orientation: 0\degree, 30\degree, 60\degree, 90\degree}
			\label{fig:19c}
		\end{minipage}
		\vspace{-0.8em}
		\caption{Impact of Orientation}
		\label{fig:19}
	\end{figure*}
	In this part, we evaluate the impact of target's orientation with respect to the radar. In the test the target's chest is towards different orientations: $\pm90\degree$, $\pm60\degree$, $\pm30\degree$ and $0\degree$, but target self is in same distance (shown in Fig. \ref{fig:19c}). During the test, the target is allowed to move their body parts. As shown in Fig. \ref{fig:19a} and \ref{fig:19b}, with the absolute degree of orientation going up, both RR and HR errors will grow. At nearly $\pm90\degree$, the radar receives little reflections directly from the chest, and the system fails to detect vital signs. The results reveal that in ideal case ($0\degree$) the target chest is directly facing the radar and the doppler caused by vital signs is totally radical and most salient thereby it leads to the lowest errors. Conversely, when the angle starts increasing till $90\degree$, the radical components of doppler will decline and finally become $0$ at $90\degree$ (the radical motion is converted to tangential motion). Consequently, vital signs doppler is no longer captured and the SNR is always 0.
	
	\subsection{Moving Targets Evaluation and Comparison}
	\begin{figure*}[h]
		\centering
		\begin{minipage}[t]{0.325\linewidth}
			\centering
			\includegraphics[width=0.93\textwidth]{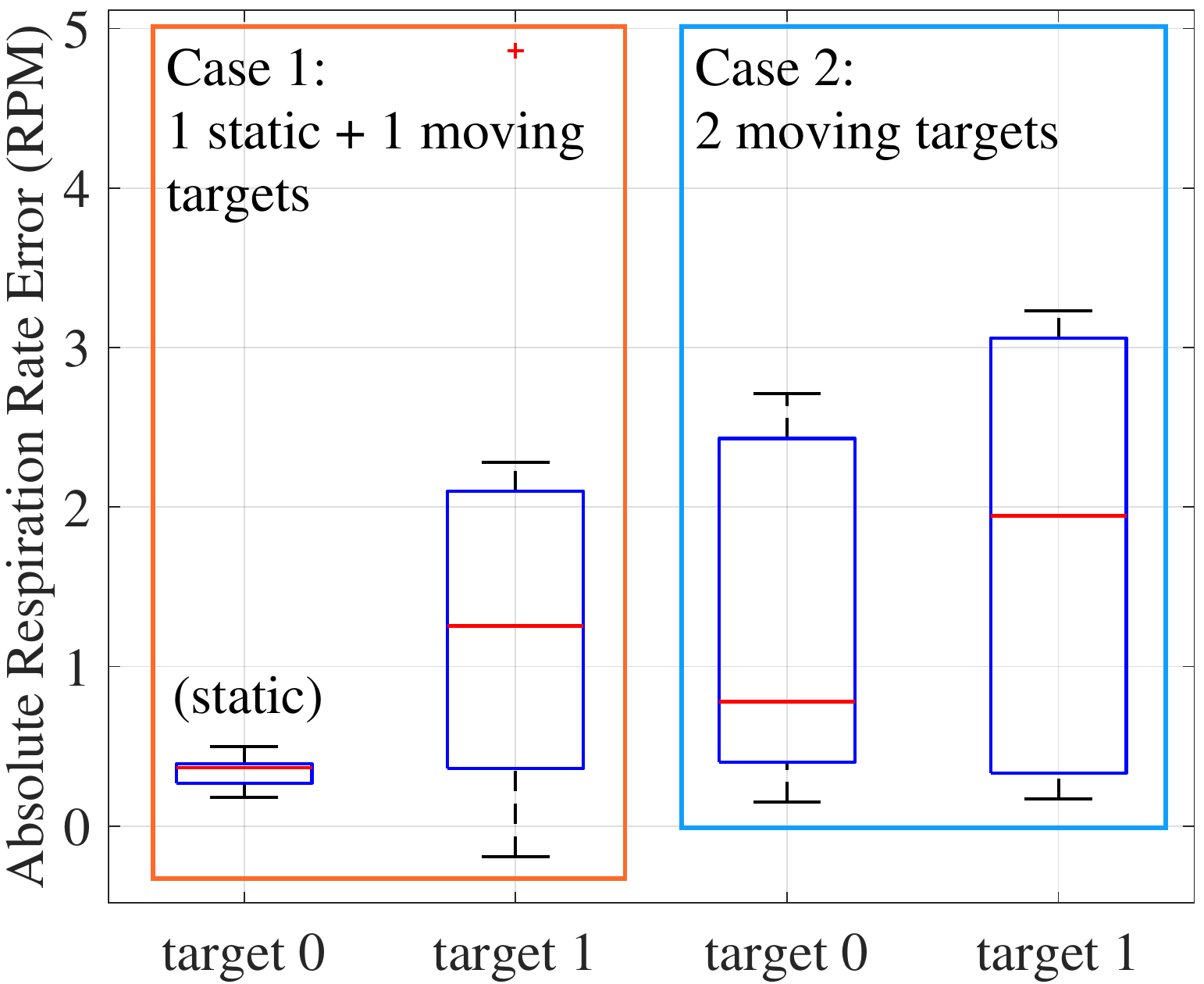}
			\subcaption{Respiration}
			\label{fig:20a}
		\end{minipage}
		\begin{minipage}[t]{0.325\linewidth}
			\centering
			\includegraphics[width=0.95\textwidth]{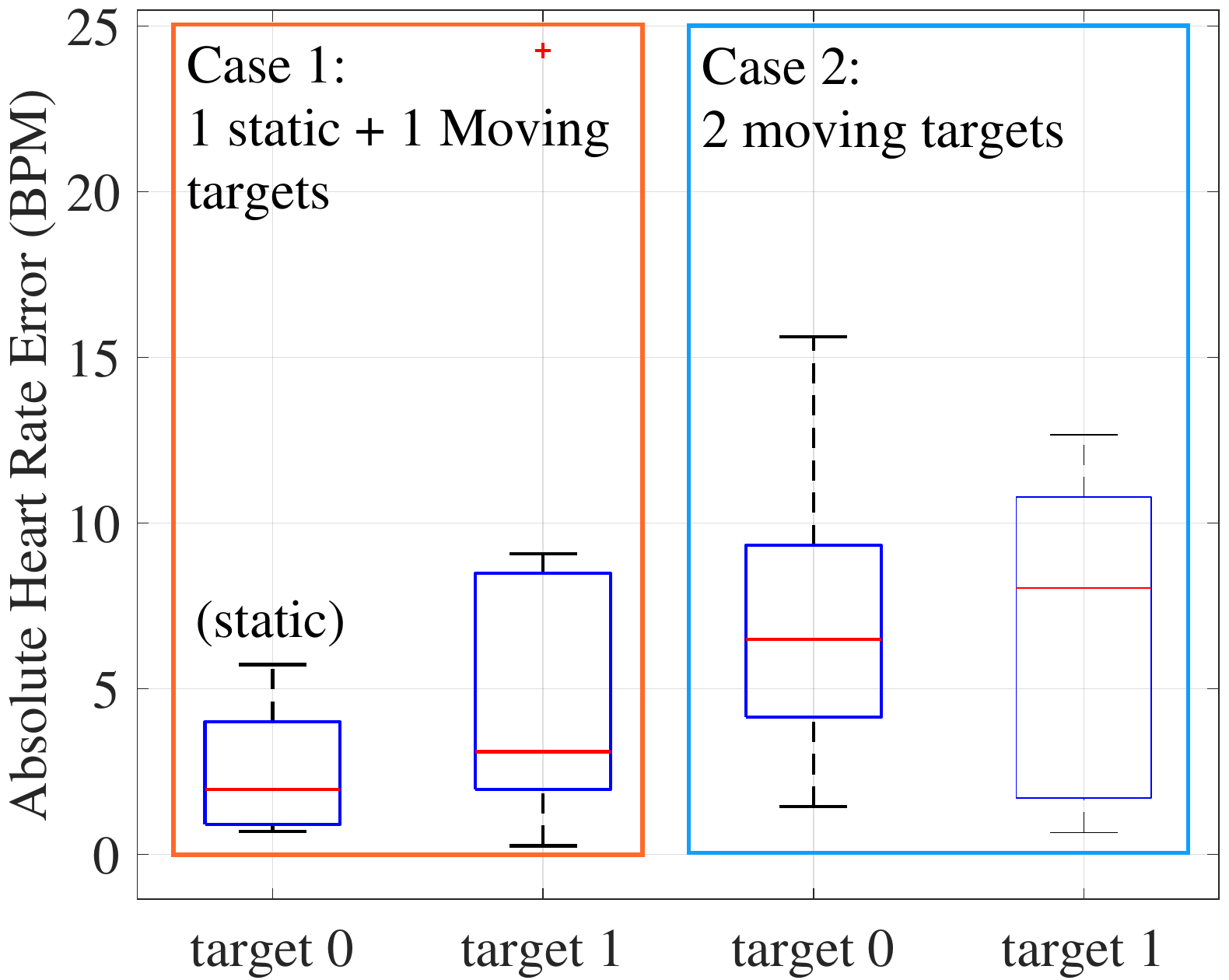}
			\subcaption{Heart}
			\label{fig:20b}
		\end{minipage}
		\begin{minipage}[t]{0.34\linewidth}
			\centering
			\includegraphics[width=1.1\textwidth, height=3.5cm]{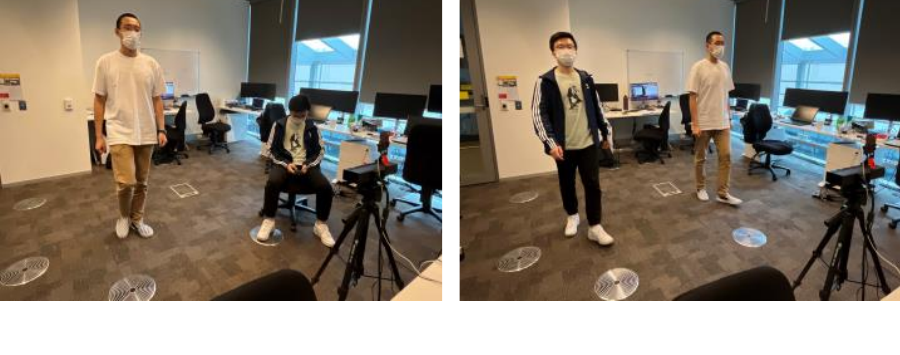}
			\subcaption{case 1: 1 static + 1 moving targets (Left)\newline case2: 2 moving targets (right) }
			\label{fig:20c}
		\end{minipage}
		\vspace{-0.8em}
		\caption{Evaluation of Moving Targets. The moving persons walk from back wall to the radar repetitively, their routines are allowed to be straight line, slash and curve.}
		\label{fig:20}
	\end{figure*}
	Although our scheme is designed and optimized for location fixed targets, it is also capable of sensing moving targets at a medium accuracy, primarily credited to the noise suppression and motion-tolerant techniques as presented in Section 6 and 7. To demonstrate the capability, we perform two experiments: (1) one location-fixed target + one moving target and (2) Two moving targets (shown in Fig. \ref{fig:20c}). In the experiments, the targets are walking from one side to another in the room and repeating this process. It should be mentioned that during target walking the orientation of targets' chest is within the range from $-60$ degree to $+60$ degree towards the radar.

	The results are presented in Fig. \ref{fig:20a} and \ref{fig:20b}, which indicates that our scheme still achieve medium sensing accuracy, which is only slightly degraded compared to the case of a location-fixed target. Overall, the errors in Experiment 1 are better than Experiment 2. For the location-fixed target, the sensing has the lowest RR and HR errors (less than 0.4 RPM and 2 BPM). This reveals that the proposed scheme is capable of separating the location-fixed and moving targets and suppressing the noise from moving one. In Experiment 1, although the target is moving, the median errors for RR and HR are still below 1.5 RPM and 5 BPM. This indicates that the proposed scheme can suppress the interference of position varying and body part motion induced by a moving target. And it is efficient in estimating vital rates for both targets. 
	
	In another experiment, when all targets are moving, the errors (highest errors: 1.95 RPM and 8.04 BPM, lowest errors: 0.78 RPM and 6.5 BPM) are higher than those in Experiment 1. A larger error is observed to often happen when the two targets move closely to each other. In this case, beamforming becomes less efficient in separating signals from the two users, which causes larger interferences to each other.

	We also compare our scheme with two state-of-the-arts for sensing moving target's respiration: BreathCatcher\cite{mov_brth_cmp} and heartbeat monitoring \cite{mov_hrt_cmp}. As shown in Table \ref{table_cmp_mov}, although BreathCatcher \cite{mov_brth_cmp} achieves less than $\frac{1}{3}$ error of ours, this data-driven scheme requires over 80 hours respiration data to train the deep learning model and reconstruct the fine-grained breath waveform in prior.
	
	More importantly, our scheme is primarily developed for sensing location-fixed targets thereby no processing or optimization is made for object-tracking and moving dynamics. On the contrary, in both \cite{mov_brth_cmp} and \cite{mov_hrt_cmp} the object tracking algorithm of Kalman filter (KF) is leveraged to obtain more accurate target position. Another important difference is that both \cite{mov_brth_cmp} and \cite{mov_hrt_cmp} use two radars while only one radar used in our scheme.
	
	\begin{table}
		\caption{Overall comparison of moving target errors }
		\vspace{-1.5em}
		\label{table_cmp_mov}
		\center
		\begin{tabular}{c|c|c|c}
			\hline
			Methods & RR (RPM)  & HR (BPM)  & Data-driven\\ \hline
			\cite{mov_brth_cmp}  & 0.4 & - & Yes\\ \hline
			\cite{mov_hrt_cmp}     & -  & 8 & No\\ \hline
			This work & 1.32 & 5.88 & No\\ \hline
		\end{tabular}
		\vspace{-1em}
	\end{table}
	
	\subsection{Multiple Targets Evaluation and Comparison}
\begin{figure*}[h]
		\centering
		\begin{minipage}[t]{0.325\linewidth}
			\centering
			\includegraphics[width=0.98\textwidth]{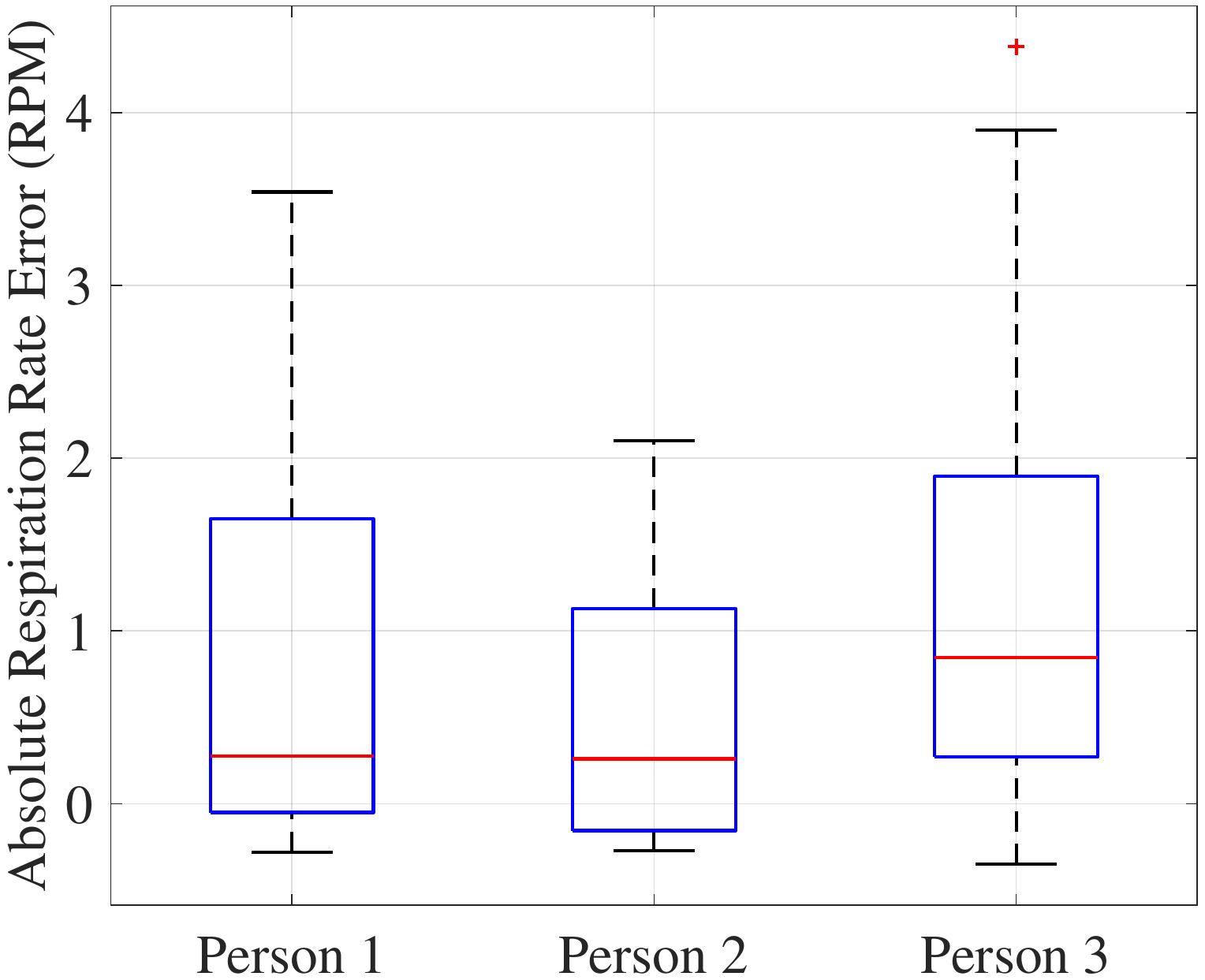}
			\subcaption{Respiration}
			\label{fig:21a}
		\end{minipage}
		\begin{minipage}[t]{0.325\linewidth}
			\centering
			\includegraphics[width=1\textwidth]{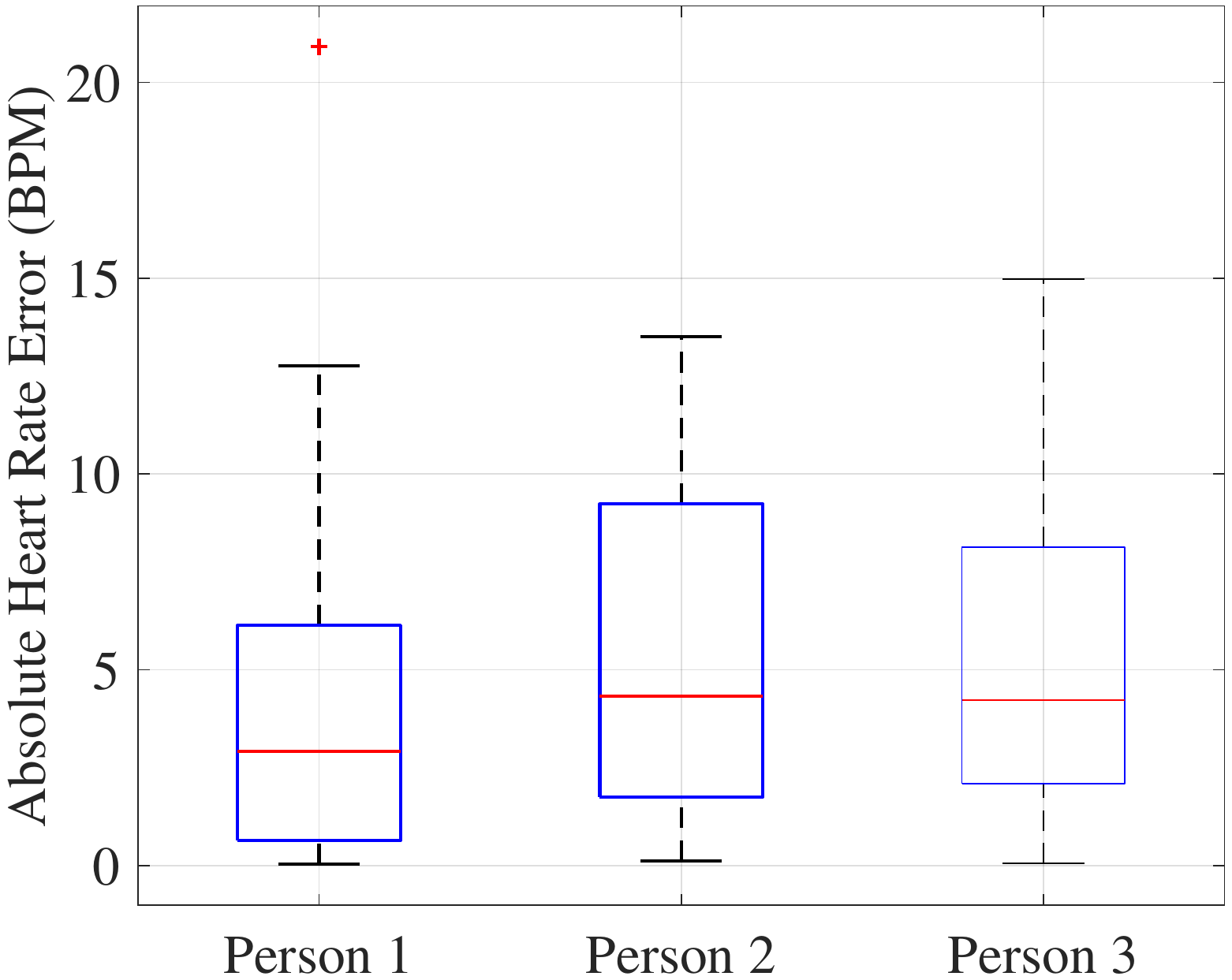}
			\subcaption{Heart}
			\label{fig:21b}
		\end{minipage}
		\begin{minipage}[t]{0.34\linewidth}
			\centering
			\includegraphics[width=0.9\textwidth, height=4.75cm]{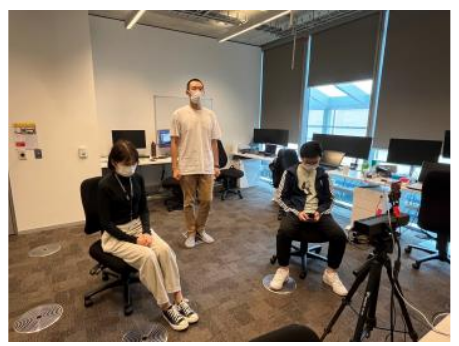}
			\subcaption{Multiple targets scene}
			\label{fig:21c}
		\end{minipage}
		\vspace{-0.8em}
		\caption{Evaluation of Multiple Targets. All the individuals are allowed to move their body motions. In this scene, the standing one is marching on spot, other two are playing with cellphone and performing freely motion.}
		\label{fig:21}
	\end{figure*}

	To further evaluate the performance in sensing multiple location-fixed targets, three participants are asked to stand or sit in front of the radar at different angles and distances (shown in Fig. \ref{fig:21c}). All the individuals are allowed to move their body parts freely and randomly. From the results, it can be observed that all median errors are similar and do not vary too much. The lowest RR and HR median errors are respectively 0.26 RPM and 2.92 BPM, which are close to the single target results. The highest errors reach 0.84 RPM for RR and 4.33 for HR. 

	We also compare our results with another two state-of-the-arts \cite{c33} \cite{cwt_hr} for sensing multiple location-fixed targets. In \cite{c33}, the phased-MIMO is employed to perform beamforming for RR and HR estimation and \cite{cwt_hr} utilizes adapted wavelet for HR estimation. The overall error comparisons are shown in Table \ref{table_cmp_multi}. It can be observed that our proposed scheme achieves the lowest RR error (0.21 RPM), supporting the sensing of three users. Despite that the work \cite{cwt_hr} has a higher accuracy in HR estimation, the radar sensor applied in \cite{cwt_hr} contains 192 channels, which can provide over 20 times higher resolution than ours (4 channels) and \cite{c33} (8 channels). Such high-resolution radar sensor can provide more powerful beamforming than ours but its price of this device is over 6 times higher. More importantly, both \cite{c33} and \cite{cwt_hr} are not motion-tolerant as their algorithms are aimed to estimate the fixed bandwidth frequency.
	
	\begin{table}
		\caption{Overall comparison of multi-target errors }
		\vspace{-1.5em}
		\label{table_cmp_multi}
		\center
		\begin{tabular}{c|c|c|c|c}
			\hline
			Methods & RR  & HR  & N of tars & Motion-tolerant\\ \hline
			\cite{c33}     & 1.06  & 4.34  &  2 & No\\ \hline
			\cite{cwt_hr}  & - & 2 & 3 & No\\ \hline
			This work & 0.21 & 3.62 & 3 & Yes\\ \hline
		\end{tabular}
		\vspace{-1em}
	\end{table}
	
	\section{Conclusion}
	This paper introduces a reliable contact-free vital sign monitoring system for heartbeat and breathing frequency estimation in the presence of static reflectors, target body motion and moving people. The key designs of the system are fusing a video camera and mmWave radar to identify the target, using beamforming to suppress interference and separating the vital sign from multiple range bins, and introducing novel algorithms for extracting weak vital-sign signals. The novel algorithm of WMC-VMD and its acceleration strategy are proposed to achieve high-precision and real-time monitoring. Our system works in different and practical scenarios without any environment-specific training. Considering the potential privacy issue with the use of a camera, our scheme is mainly suitable for such as health monitoring in nursing homes, in-home elder healthcare and inmates-care in prison, where the privacy is secured. For more privacy-focused scenarios, our scheme can be readily adapted to the case when infrared or low-resolution camera is used.
	
	Our future work is to monitor the vital signs of dynamic targets. Object tracking methods are required to accurately locate the targets. Once the accurate target positions are obtained, advanced learnable algorithms can be further used to achieve vital sign recognition, by suppressing the impact of large body movement. In addition, the usage of camera in our current scheme limits the applications to visually LoS sensing; it is also an important future task to extend the scheme to achieve NLoS sensing with a pure radar solution. This will be based on innovative radar techniques for picking up targets of interest in the presence of both moving interferers and background clutter.


	\ifCLASSOPTIONcompsoc


	\ifCLASSOPTIONcaptionsoff
	\newpage
	\fi

	\bibliographystyle{IEEEtran}
	\bibliography{bibliography_2}

\end{document}